\documentclass[DIV=12,numbers=noenddot]{scrartcl}

\usepackage{ifluatex}
\ifluatex
\else
\usepackage[utf8]{inputenc}
\usepackage[T1]{fontenc}
\fi

\usepackage{lmodern}
\usepackage{amsmath,amssymb}
\usepackage{mathtools}
\usepackage{xcolor}
\usepackage{scalefnt}
\usepackage{xspace}
\usepackage[final]{listings}
\usepackage{booktabs}
\usepackage{authblk}

\usepackage[font=small,labelfont=bf,format=plain,margin=0.05\textwidth]{caption}
\usepackage{subcaption}
\usepackage{tabularx}

\newcommand{\mytitle}{From Lagrangian to Higgs physics constraints for SUSY and non-SUSY models: interfacing FlexibleSUSY with HiggsTools and Lilith}

\usepackage[
  pdftitle={\mytitle},
  pdfauthor={Wojciech Kotlarski, Alexander Voigt},
  pdfkeywords={FlexibleSUSY, Higgs boson, HiggsTools, Lilith},
  bookmarks=true, linktocpage, colorlinks=true, allbordercolors=white,
  allcolors=blue]{hyperref}
\usepackage{orcidlink} 

\usepackage[
  backend=biber,
  bibencoding=utf8,
  sorting=none,
  url=true,
  doi=false,
  style=phys,
  biblabel=brackets,
  subentry=false,
  eprint=true,
  maxbibnames=5
]{biblatex}
\title{\mytitle}
\date{}
\author[a]{Wojciech Kotlarski\orcidlink{0000-0002-1191-6343}\thanks{\href{mailto:wojciech.kotlarski@ncbj.gov.pl}{wojciech.kotlarski@ncbj.gov.pl}}}
\affil[a]{National Centre for Nuclear Research, Pasteura 7, 02-093 Warsaw, Poland}
\author[b]{Alexander Voigt\orcidlink{0000-0001-8963-6512}\thanks{\href{mailto:alexander.voigt@physik.rwth-aachen.de}{alexander.voigt@physik.rwth-aachen.de}}}
\affil[b]{Institute for Theoretical Solid State Physics, RWTH Aachen University, Sommerfeldstra{\ss}e 16, 52074 Aachen, Germany}

\newcommand{\abbrev}[1]{\text{#1}} 

\newcommand{\CO}{\texttt{COLLIER}\@\xspace}
\newcommand{\FD}{\texttt{Flex\-ib\-le\-De\-cay}\@\xspace}
\newcommand{\FS}{\texttt{Flex\-ib\-le\-SUSY}\@\xspace}
\newcommand{\HT}{\texttt{Higgs\-Tools}\@\xspace}
\newcommand{\HS}{\texttt{Higgs\-Signals}\@\xspace}
\newcommand{\HB}{\texttt{Higgs\-Bounds}\@\xspace}
\newcommand{\LL}{\texttt{Lilith}\@\xspace}
\newcommand{\LT}{\texttt{Loop\-Tools}\@\xspace}
\newcommand{\PY}{\texttt{python}\@\xspace}
\newcommand{\SA}{\texttt{SARAH}\@\xspace}
\newcommand{\SLHA}{\texttt{SLHA}\@\xspace}
\newcommand{\SO}{\texttt{SOFT\-SUSY}\@\xspace}
\newcommand{\WL}{\texttt{Wolfram Language}\@\xspace}

\newcommand{\BR}{\ensuremath{\text{BR}}}
\newcommand{\colvec}[1]{\begin{pmatrix}#1\end{pmatrix}}
\newcommand{\ee}{\text{e}}
\newcommand{\SM}{\ensuremath{\abbrev{SM}}\xspace}
\newcommand{\BSM}{\ensuremath{\abbrev{BSM}}\xspace}

\newcommand{\MSSM}{\ensuremath{\abbrev{MSSM}}\xspace}
\newcommand{\NMSSM}{\ensuremath{\abbrev{NMSSM}}\xspace}

\renewcommand{\imath}{\ensuremath{\text{i}}}

\newcommand{\sectionname}{Section}
\newcommand{\secref}[1]{\sectionname~\ref{#1}}
\newcommand{\figref}[1]{\figurename~\ref{#1}}
\newcommand{\tabref}[1]{\tablename~\ref{#1}}
\newcommand{\code}[1]{\lstinline|#1|} 

\newcommand{\TeV}{\,\text{TeV}}
\newcommand{\GeV}{\,\text{GeV}}
\newcommand{\MeV}{\,\text{MeV}}

\newcommand{\smc}[1]{\hat{#1}}   

\newcommand{\CPP}{C\nolinebreak\hspace{-.05em}\raisebox{.4ex}{\tiny\textbf +}\nolinebreak\hspace{-.10em}\raisebox{.4ex}{\tiny\textbf +}\@\xspace}

\begin{document}
\maketitle
\begin{abstract}
  \FS is a framework for an automated calculation of observables in user-defined models of a Beyond the Standard Model (BSM) physics, starting from the model's field content and its Lagrangian.
  Among a plethora of observables it is capable of calculating are the high precision predictions for Higgs bosons decay widths.
  Building on these previous developments we present here an interface between \FS and \HT/\LL.
  Combined with other \FS capabilities this extension provides a fully automatized tool chain leading directly from a user-defined BSM model to the state-of-the-art validation of the global agreement of a BSM Higgs sector with experimental measurements.
  We demonstrate this extension on a handful of phenomenologically relevant examples: a CP-conserving version of the Type-II Two Higgs Doublet Model, the CP-violating Next-to-Minimal Supersymmetric Standard Model and the Minimal R-symmetric Supersymmetric Standard Model.
  These examples show the power of \FS when applied to supersymmetric and non-supersymmetric models, both with and without CP-violation, and illustrate the handling of invisible and undetected decay widths.
\end{abstract}

\tableofcontents

\section{Introduction}

One of the main goals of future collider experiments is a more precise measurement of Higgs boson properties.
High Luminosity LHC, FCC-$hh$ and considered $e^+e^-$ colliders like CLIC, ILC, FCC-$ee$, CEPC and C$^3$ have planed an extensive Higgs physics program \cite{Cepeda:2019klc,FCC:2018byv,Bambade:2019fyw}, with a dedicated Higgs factory identified by the European Strategy for Particle Physics as one of its priorities \cite{EuropeanStrategyGroup:2020pow}.
Apart from SM-like Higgs measurements there are also ongoing preparations for a search of Beyond the Standard Model (BSM) scalars, with for example the search of light Higgs-like scalar in the ``Higgs-strahlung'' process being defined as one of focus topics for the ECFA study on Higgs/Top/EW factories \cite{deBlas:2024bmz}.

Irrespectively of future experiments, currently available data already now puts strong constrains on any BSM theory.
This is however conditioned on the assumption that we are able to reliably predict properties of BSM Higgs bosons in a given model and compare them with experimental data, ideally in an automatized fashion.
To that end we have created an automatized way to check the validity of BSM Higgs sectors using the programs \HT \cite{Bahl:2022igd} and \LL \cite{Bernon:2015hsa,Kraml:2019sis,Bertrand:2020lyb} within the framework of \FS, which we present in this work.

\FS \cite{Athron:2014yba,Athron:2017fvs} is a \WL/\CPP computer program created to facilitate studies of phenomenology of BSM models, starting from the definition of the fields and their interactions.
It is based on \SA \cite{Staub:2013tta,Staub:2009bi,Staub:2010jh,Staub:2012pb} and includes components of \SO \cite{Allanach:2001kg,Allanach:2013kza}.
It automatically computes a plethora of phenomenologically relevant observables, like loop-corrected mass spectra, in particular Higgs boson masses \cite{Athron:2016fuq,Kwasnitza:2020wli,Kwasnitza:2025mge}, low-energy observables like lepton $g-2$ and electric dipole moments \cite{Athron:2017fvs}, $b$-physics observables \cite{Khasianevich:2024hpv}, $W$-boson mass \cite{Athron:2022isz}, scalar decays \cite{Athron:2021kve} and more.
\FS can calculate these observables in a broad class of user-defined supersymmetric (SUSY) and non-supersymmetric (non-SUSY) models.
For example, it has been successfully applied in studies of muon $g-2$ and charged lepton flavour violation (cLFV) in the Minimal R-symmetric Supersymmetric Standard Model (MRSSM) \cite{Kotlarski:2019muo} and in a scalar leptoquark model \cite{Dudenas:2022von}, as well as in the study of neutrino masses and cLFV in the Grimus-Neufeld model \cite{Dudenas:2022von,Dudenas:2022xnq}, and in a general study of muon $g-2$ in a broad collection of SUSY and non-SUSY models \cite{Athron:2021iuf} and many others.

A recently developed decay module of \FS, called \FD \cite{Athron:2021kve}, allows to compute branching ratios of scalar fields, with a special emphasis on precise calculation of Higgs boson decays.
Until now, \FD provided only total widths and branching ratios.
This made it cumbersome to quickly asses whether a given parameter point is or is not excluded by currently available experimental data in a statistically meaningful way.
For this purpose however, dedicated computer programs \HT and \LL where created.

In this work we present a \FS interface to \HT and \LL which allows for a seamless determination, whether a given parameter point of a BSM model is compatible with the Higgs data.
The procedure is fully automatized and does not require any intervention from the user.
This extension has already been used in phenomenological analyses, for instance in the explanation of a potential $95\GeV$ excess seen at LEP and LHC through a Higgs-like state from the MRSSM \cite{Kalinowski:2024uxe}, and in the study of the Type-II Two Higgs Double Model (2HDM) parameter points originating from the principle of Reduction of Couplings \cite{Kotlarski:2025jvf}.
This extension has already been used in phenomenological analyses, for instance in the explanation of a potential $95\GeV$ excess seen at LEP and LHC as originating from a Higgs-like state from the MRSSM \cite{Kalinowski:2024uxe}, and in the study of the Type-II Two Higgs Double Model (2HDM) parameter points originating from the principle of Reduction of Couplings \cite{Kotlarski:2025jvf}.
In this paper, we will describe the details of this interface.

This paper is structured as follows:
In \secref{sec:quick} we describe how to install \FS in combination with \HT and/or \LL.
In \secref{sec:implementation} we describe the design of the interface between \FS, \HT and \LL.
In \secref{sec:application} we show example applications in the Type-II 2HDM, the CP-violating Next-to-Minimal Supersymmetric Standard Model (NMSSM) and in the MRSSM.
We summarize and conclude in \secref{sec:summary}.

\section{Quick start guide}
\label{sec:quick}

The \HT/\LL interface requires \FS~2.9.0, which can be downloaded from
the \FS github repository
[\url{https://github.com/FlexibleSUSY/FlexibleSUSY}] by running
\begin{lstlisting}[language=bash]
$ wget https://github.com/FlexibleSUSY/FlexibleSUSY/archive/refs/tags/v2.9.0.tar.gz -O FlexibleSUSY-2.9.0.tar.gz
$ tar -xf FlexibleSUSY-2.9.0.tar.gz
$ cd FlexibleSUSY-2.9.0
\end{lstlisting}
All prerequisites listed in the \FD manual \cite{Athron:2021kve} are
required.  In particular \FS has to be configured with a dedicated
library to calculate 1-loop functions, where currently \LT
\cite{Hahn:1998yk} and \CO
\cite{Denner:2002ii,Denner:2005nn,Denner:2010tr,Denner:2016kdg} are
supported.  Additionally, \HT~1.1.4 or later and/or \LL\footnote{At
  the time of writing, the latest version \LL~2.1 is incompatible with
  \PY~3. However, a fix is already available in the master branch of
  \LL's development version. Thus, in order to use \LL with \FS, \LL
  has to be build from the master branch.} must be installed. We refer
the reader to respective manuals for installation instructions.  \HT
datasets are distributed separately from its source code and have to
be downloaded separately.  The \LL interface requires \PY~3
development files, which may be installed via a local package manager.
On Ubuntu, the appropriate package is called \texttt{python3-dev}, for
example.

We assume that \lstinline|CO_DIR|, \lstinline|LL_DIR|,
\lstinline|LT_DIR| and \lstinline|HT_DIR| contain the paths to the
\CO, \LL, \LT and \HT\ directories, respectively. A \FS\ spectrum
generator for the Scalar Singlet Model \texttt{SSM} can then be build
by running
\begin{lstlisting}[language=bash]
$ ./createmodel --name=SSM -f
$ ./configure \
    --with-models=SSM \
    --with-loop-libraries=collier,looptools \
    --with-collier-incdir=$CO_DIR/modules \
    --with-collier-libdir=$CO_DIR \
    --with-looptools-incdir=$LT_DIR/build \
    --with-looptools-libdir=$LT_DIR/build \
    --with-higgstools-libdir=$HT_DIR/install/lib \
    --with-higgstools-incdir=$HT_DIR/install/include \
    --with-lilith=$LL_DIR
$ make
\end{lstlisting}
After the build has finished, the spectrum generator can be run on the
command line or from a \WL\ interface. Both interfaces are explained
in the next subsections.

\subsection{Command-line interface}
\label{sec:cmd}

To run \FS's spectrum generator for the \texttt{SSM} together with \HT
on the command line, we assume that \lstinline|HB_DATA_DIR| and
\lstinline|HS_DATA_DIR| contain the paths to the \HB\ and \HS\
datasets, respectively. We use the default \SLHA\ input file
\lstinline|models/SSM/LesHouches.in.SSM| for the \texttt{SSM}, shipped
with \FS\ \cite{Skands:2003cj,Allanach:2008qq}. The default \SLHA\
file requires some modification, though: We have to specify the loop
library by setting \lstinline|FlexibleSUSY[31]=1| (for \CO) or
\lstinline|2| (for \LT). In addition, we have to enable \HT\ by
setting \lstinline|FlexibleDecay[7]=1|. Finally, we modify the model
parameters such that $M_h\approx 125\GeV$:
\begin{lstlisting}[language=bash]
$ { cat models/SSM/LesHouches.in.SSM
    cat <<EOF
Block FlexibleSUSY
   31   1      # loop library (1 = COLLIER, 2 = LoopTools)
Block FlexibleDecay
    7   1      # call HiggsTools
Block EXTPAR   # Input parameters
    0   1000   # input scale Qin
    1   173.34 # EWSB scale QEWSB
    2   0.21   # Lambda(Qin)
    3   0.1    # LambdaS(Qin)
    4   100    # Kappa(Qin)
    5  -100    # K1(Qin)
    6   0.1    # K2(Qin)
    7   3.0    # vS(QEWSB)
EOF
} | models/SSM/run_SSM.x --slha-input-file=- \
  --higgsbounds-dataset=$HB_DATA_DIR --higgssignals-dataset=$HS_DATA_DIR
\end{lstlisting} 
The command-line output contains the following \HT\ output blocks,
which are described in \secref{sec:ht_implementation}:
\begin{lstlisting}
Block HIGGSSIGNALS
     1     1.59000000E+02   # number of degrees of freedom
     2     1.51419029E+02   # chi^2
     3     1.52548868E+02   # SM chi^2 for mh = 125.090000 GeV
     4     1.00000000E+00   # p-value
Block HIGGSBOUNDS
 25  1     7.59754148E-01   # LHC8 [...]
 25  2     3.82205822E+00   # expRatio
\end{lstlisting}
To enable \LL, we need to set \lstinline|FlexibleDecay[8]=1| and pass
the path to the dataset to the spectrum generator:
\begin{lstlisting}[language=bash]
$ { cat models/SSM/LesHouches.in.SSM
    cat <<EOF
Block FlexibleSUSY
   31   1      # loop library (1 = COLLIER, 2 = LoopTools)
Block FlexibleDecay
    8   1      # call Lilith
Block EXTPAR   # Input parameters
    0   1000   # input scale Qin
    1   173.34 # EWSB scale QEWSB
    2   0.21   # Lambda(Qin)
    3   0.1    # LambdaS(Qin)
    4   100    # Kappa(Qin)
    5  -100    # K1(Qin)
    6   0.1    # K2(Qin)
    7   3.0    # vS(QEWSB)
EOF
} | models/SSM/run_SSM.x --slha-input-file=- \
  --lilith-db=$LL_DIR/data/latestRun2.list
\end{lstlisting} 
The command-line output contains the following \LL\ block, which is
described in \secref{sec:ll_implementation}:
\begin{lstlisting}
Block LILITH
     1     5.30000000E+01   # number of degrees of freedom
     2     4.84931412E+01   # chi^2
     3     4.83975067E+01   # SM chi^2 for mh = 125.090000 GeV
     4     9.53308007E-01   # p-value
\end{lstlisting}
If no dataset for \LL is provided, the default one is used.  Note that
\HT and \LL can be enabled both at the same time.

\subsection{\WL\ interface}
\label{sec:wl_interface}

To run the \texttt{SSM} spectrum generator from \WL, one can start
from the \WL\ script \lstinline|models/SSM/run_SSM.m| generated by
\FS.  In this pre-generated script the default settings have to be
adapted, though: The calculation of the \SM\ masses must be enabled by
setting \lstinline|calculateStandardModelMasses->1|. A loop library
must be selected via \lstinline|loopLibrary->1| (for \CO) or
\lstinline|2| (for \LT). Furthermore, the model parameters must be set
to reasonable values by adapting
\lstinline|fsModelParameters|. Finally, \HT\ and \LL\ must be enabled
in \lstinline|fdSettings|. A minimal example script could look as
follows:
\begin{lstlisting}
Get["models/SSM/SSM_librarylink.m"];

handle = FSSSMOpenHandle[
    fsSettings -> {
        calculateStandardModelMasses -> 1,
        loopLibrary -> 1
    },
    fsModelParameters -> {
        Qin -> 1000,
        QEWSB -> 173.34,
        Lambdainput -> 0.21,
        LambdaSinput -> 0.1,
        Kappainput -> 100,
        K1input -> -100,
        K2input -> 0.1,
        vSInput -> 3
    },
    fdSettings -> {
        callHiggsTools -> 1,
        callLilith -> 1
    }
];

hbDatasetPath = "hbdataset-v1.7";
hsDatasetPath = "hsdataset-v1.1";
llDatasetFile = FileNameJoin[{"Lilith-2", "data", "latestRun2.list"}];

spectrum   = FSSSMCalculateSpectrum[handle];
decays     = FSSSMCalculateDecays[handle];
couplings  = FSSSMCalculateNormalizedEffectiveCouplings[handle];
higgstools = FSSSMCallHiggsTools[handle, hsDatasetPath, hbDatasetPath];
lilith     = FSSSMCallLilith[handle, llDatasetFile];

FSSSMCloseHandle[handle];
\end{lstlisting}
The function \lstinline|FSSSMOpenHandle| creates a new handle to the
parameter point, which must be closed at the end by calling
\lstinline|FSSSMCloseHandle|. The functions
\lstinline|FSSSMCalculateSpectrum|, \lstinline|FSSSMCalculateDecays|
and \lstinline|FSSSMCalculateNormalizedEffectiveCouplings| calculate
the mass spectrum, the decays and the normalized effective couplings,
respectively. The function \lstinline|FSSSMCallHiggsTools| calls \HT\
with the given paths to the \HS\ and \HB\ datasets passed as second
and third argument. Similarly, the function
\lstinline|FSSSMCallLilith| calls \LL\ with the given dataset file
passed as second argument.\footnote{While the code to call \LL from
  \FS via \texttt{Mathematica}/\texttt{Wolfram Engine} has been
  prepared, currently there is an issue in the internall \LL call to
  \texttt{NumPy} \cite{harris2020array} when called via the \CPP
  bridge.  For the time being \LL will therefore be accessible only
  via the command-line interface (as described in \secref{sec:cmd}).}
For the parameter point given in the \WL\ script above the content of
the variable \lstinline|higgstools| reads
\begin{lstlisting}
{SSM -> {
  {ndof -> 159, chi2 -> 151.514, chi2SM -> 152.549, pval -> 1.}, 
  {{ pdgid -> 25, obsRatio -> 0.759372, expRatio -> 3.82014, 
     "bestAnalysis" -> "LHC8 [vbfH,HW,Htt,H,HZ]>[bb,tautau,WW,ZZ,gamgam] \
       from CMS-PAS-HIG-12-045 (CMS 17.3fb-1, M=(110, 600))" }}
}}
\end{lstlisting}
The output mirrors the content of the \HS and \HB \SLHA\ output blocks.

\section{Implementation}
\label{sec:implementation}

The properties of the BSM Higgs boson $\phi$ in relation to SM particles can be described by the following ad-hoc phenomenological Lagrangian
\begin{align}
\label{eq:effLag}
 \mathcal{L}^\text{CP-even}_{\phi f \bar{f}} ={}& c_f \frac{\sqrt{2} m_f}{v} \bar{f} f \phi + c_W \frac{2 m_W^2}{v} W_\mu^+ W^{-\mu} \phi + c_Z \frac{m_Z^2}{v} Z_\mu Z^\mu \phi \nonumber \\
 & + c_g G_{\mu \nu} G^{\mu \nu} \phi + c_\gamma F_{\mu \nu} F^{\mu \nu} \phi + c_{\gamma Z} Z_{\mu \nu} \tilde{F}^{\mu \nu} \phi, \\
 \mathcal{L}^\text{CP-odd}_{\phi f \bar{f}} ={}& \imath \tilde{c}_f \frac{\sqrt{2} m_f}{v} \bar{f} \gamma^5 f \phi,
\end{align}
where $\tilde{F}$ is the dual electromagnetic field strength tensor, $\tilde{F}^{\mu\nu} = \epsilon^{\mu \nu \rho \sigma} F_{\rho\sigma}/2$ (and analogously for the gluon tensor $G$), and where the effective CP-even couplings $c$ and CP-odd couplings $\tilde{c}$ parametrize the BSM nature of the Higgs interactions with SM particles.
The normalization is chosen such that at the tree-level the SM predicts $c_f=c_Z=c_W=1$ and $c_g=c_\gamma=c_{\gamma Z}=0$.
Moreover, in models without CP-violation, $\tilde{c}_i \equiv 0$.

The effective couplings $c$ and $\tilde{c}$ can be computed from Higgs partial widths with \FS, including  higher-order corrections, since the introduction of the \FD \cite{Athron:2021kve}.
By computing them from loop-corrected partial widths, as opposed to tree-level vertices, one hopes to capture some of the higher-order BSM effects.
Currently, \FS includes mostly higher-order SM corrections, with only a limited capability to include pure BSM contributions.
This approach is adopted throughout the literature for loop-induced couplings.

As an example, in this approach the effective coupling to $Z$ boson, $c_Z$, is extracted from the $\Gamma(\phi \to ZZ)$ partial width as
\begin{equation}
c_Z^2 \sim \Gamma(\phi \to Z Z).
\end{equation}
This approach neglects potential dimension-5 contributions which are sensitive to the CP properties of the Higgs. (Note that the lowest order dimension-3 operator is always CP even.)
This approximation is also what is implemented in \HT nor \LL.

The fermion couplings $c_f$ and $\tilde{c}_f$ can be similarly extracted from the partial width, which has the form
\begin{equation}
\label{eq:fermion_partial}
\Gamma(\phi \to f \bar{f}) \sim c_f^2 \beta^3 + \tilde{c}_f^2 \beta,
\end{equation}
where $\beta$ is the fermion velocity (assuming here for illustration that both fermions have the same mass).
In this case we do keep track of the CP-structure of the coupling, distinguishing between $c_f$ and $\tilde{c}_f$.

The only exception to the extraction of couplings from widths is the Higgs coupling to top-quarks.
This decay is kinematically forbidden in case of the SM-like Higgs.
If that is the case, we extract $c_t$ and $\tilde{c}_t$ from the tree-level vertex.

Since all of those couplings, up to their CP-structure, exist also in the SM, it is convenient to normalize them to an equivalent coupling for a SM Higgs of mass equal to the $\phi$ mass.
This gives the so-called $\kappa$ framework, where\footnote{The $\kappa$ framework is often expressed purely in terms of partial widths. This, however, looses track of the CP-structure of interactions.}
\begin{equation}
\label{eq:kappa_def}
\kappa_i = c_i/c_i^\SM, \qquad \tilde{\kappa}_i = \tilde{c}_i/c_i^\SM.
\end{equation}
A pure SM corresponds then to $\kappa_i = 1$ and $\tilde{\kappa}_i = 0$, also in the case of loop-induced couplings. 

\subsection{Construction of (normalized) effective Higgs couplings}

The construction of normalized effective couplings in \FS proceeds as follows.
For every BSM model we construct a corresponding SM with equivalent settings and compute decays in both models.
The \FS settings relevant to this discussion are summarized in \tabref{tab:FS_SLHA_configuration_FD-block}.
The options 0--4 control the calculation of decays and were already described in the \FD manual \cite{Athron:2021kve}.
The calculation of reduced effective couplings is controlled by flag~6.
The new feature, present in \FS~2.9.0 and first used in Ref.~\cite{Kalinowski:2024uxe}, is the use of pole scalar, pseudoscalar and charged Higgs mixing matrices in the vertices that enter the decay calculation.
This is the current default.

\begin{table}
  \centering
  \begin{tabularx}{\textwidth}{clcX}
    \toprule
    Index & mathematica symbol & Default & Description \\
    \midrule
    0 & -- & $1$ & calculate decays (0 = no, 1 = yes) \\
    1 & \texttt{minBRtoPrint} & $10^{-5}$ & minimum BR to print\\
    2 & \texttt{maxHigherOrderCorrections} & 4 & include higher order corrections in decays (0 = \texttt{LO}, 1 = \texttt{NLO}, 2 = \texttt{NNLO}, 3 = \texttt{NNNLO}, 4 = \texttt{NNNNLO}) \\
    3 & \texttt{alphaThomson} & 1 & use $\smc\alpha(m)$ or Thomson $\alpha(0)$ in decays to $\gamma \gamma$ and $\gamma Z$ (0 = $\smc\alpha(m)$, 1 = $\alpha(0)$) \\
    4 & \texttt{offShellVV} & 2 & decays into off-shell $VV$ pair (0 = no off-shell decays, 1 = single off-shell decays if $m_V < m_H < 2 m_H$, double off-shell if $m_H < m_V$, 2 = double off-shell decays if $m_H < 2 m_V$) \\
    5 & \texttt{printEffC} & 1 & print loop-induced Higgs couplings for use with \SA generated \texttt{UFO} and \texttt{CalcHEP} models \\
    6 & \texttt{calcNormalizedEffC} & 0 & calculate effective Higgs couplings to SM particles normalized to their SM-like values for a SM with $m_h^\SM \equiv m_{H_i}^\BSM$ or $m_{A_i}^\BSM$\\
    7 & \texttt{callHiggsTools} & 0 & call \HT\\
    8 & \texttt{callLilith} & 0 & call \LL \\
    9 & \texttt{usePoleHiggsMixings} & 1 & use Higgs pole mixing matrices in vertices for decays \\
    \bottomrule
  \end{tabularx}
  \caption{Entries for the \FD \SLHA input block and
    corresponding \texttt{Mathematica} symbols to specify the runtime
    configuration options for \FD and its interface to \HT and \LL.
    Options 5 is new in \FS~2.8.0, options $\geq 6$ are new in 2.9.0, see \tablename~A.2 in Ref.~\cite{Athron:2021kve}.}
  \label{tab:FS_SLHA_configuration_FD-block}
\end{table}

We have performed an internal consistency check by computing
$\kappa_i$ and $\tilde{\kappa}_i$ from Eq.~\eqref{eq:kappa_def} for
the case when the BSM model is taken to be the SM.  In this case we
expect $\kappa_i=1$ and $\tilde\kappa=0$. Numerically the largest
observed deviation, $\max(|\kappa_i - 1|, |\tilde{\kappa}_i|)$, that
we find is below 0.05\%, which is below the precision of the effective
coupling approximation.

The following snippet shows an example output of the normalized
effective couplings for the Type-II 2HDM parameter point from
Eq.~\eqref{eq:2hdm_bmp} from \secref{sec:2hdm}.  The couplings
$\kappa$ and the vector boson subset of the CP-odd couplings
$\tilde{\kappa}$ are written to the
\texttt{NORMALIZEDEFFHIGGSCOUPLINGS} \SLHA block:
\begin{lstlisting}
Block NORMALIZEDEFFHIGGSCOUPLINGS
   25     0     0   4.06234679E-03 # SM Higgs width for mhSM = mhh(1)
   25    -1     1   1.00001694E+00 # hh(1)-barFu(1)-Fu(1)/SM with mhSM = mhh(1)
   25    -2     2   9.99870861E-01 # hh(1)-barFd(1)-Fd(1)/SM with mhSM = mhh(1)
   25    -3     3   9.99868935E-01 # hh(1)-barFd(2)-Fd(2)/SM with mhSM = mhh(1)
   25    -4     4   1.00000880E+00 # hh(1)-barFu(2)-Fu(2)/SM with mhSM = mhh(1)
   25    -5     5   9.99707996E-01 # hh(1)-barFd(3)-Fd(3)/SM with mhSM = mhh(1)
   25    -6     6   1.00001984E+00 # hh(1)-barFu(3)-Fu(3)/SM with mhSM = mhh(1)
   25   -11    11   9.99862385E-01 # hh(1)-barFe(1)-Fe(1)/SM with mhSM = mhh(1)
   25   -13    13   9.99862385E-01 # hh(1)-barFe(2)-Fe(2)/SM with mhSM = mhh(1)
   25   -15    15   9.99862385E-01 # hh(1)-barFe(3)-Fe(3)/SM with mhSM = mhh(1)
   25   -24    24   9.99603798E-01 # hh(1)-conjVWm-VWm/SM with mhSM = mhh(1)
   25    23    23   9.99816512E-01 # hh(1)-VZ-VZ/SM with mhSM = mhh(1)
   25    21    21   9.99948351E-01 # hh(1)-VG-VG/SM with mhSM = mhh(1)
   25    22    22   1.00209485E+00 # hh(1)-VP-VP/SM with mhSM = mhh(1)
   25    23    22   1.00063357E+00 # hh(1)-VP-VZ/SM with mhSM = mhh(1)
   ...
   36     0     0   3.99173109E+01 # SM Higgs width for mhSM = mAh(2)
   36    21    21   4.44586838E-01 # Ah(2)-VG-VG/SM with mhSM = mAh(2)
   36    22    22   1.08288778E+00 # Ah(2)-VP-VP/SM with mhSM = mAh(2)
   36    23    22   8.86773505E-02 # Ah(2)-VP-VZ/SM with mhSM = mAh(2)
\end{lstlisting}
Accordingly, the CP-odd parts of the Higgs boson couplings to
fermions, $\tilde{\kappa}_f$, are written to the
\texttt{IMNORMALIZEDEFFHIGGSCOUPLINGS} block.\footnote{We follow the
  \SLHA~2 \cite{Allanach:2008qq} convention and prefix the block name
  containing imaginary parts with \texttt{IM}.} For the 2HDM parameter
point from above, the corresponding output reads
\begin{lstlisting}
Block IMNORMALIZEDEFFHIGGSCOUPLINGS
   36    -1     1   3.21035866E-01 # Ah(2)-barFu(1)-Fu(1)/SM with mhSM = mAh(2)
   36    -2     2   3.08230893E+00 # Ah(2)-barFd(1)-Fd(1)/SM with mhSM = mAh(2)
   36    -3     3   3.11075400E+00 # Ah(2)-barFd(2)-Fd(2)/SM with mhSM = mAh(2)
   36    -4     4   3.18026781E-01 # Ah(2)-barFu(2)-Fu(2)/SM with mhSM = mAh(2)
   36    -5     5   3.13510056E+00 # Ah(2)-barFd(3)-Fd(3)/SM with mhSM = mAh(2)
   36    -6     6   2.98994917E-01 # Ah(2)-barFu(3)-Fu(3)/SM with mhSM = mAh(2)
   36   -11    11   3.14775680E+00 # Ah(2)-barFe(1)-Fe(1)/SM with mhSM = mAh(2)
   36   -13    13   3.14775680E+00 # Ah(2)-barFe(2)-Fe(2)/SM with mhSM = mAh(2)
   36   -15    15   3.14775714E+00 # Ah(2)-barFe(3)-Fe(3)/SM with mhSM = mAh(2)
\end{lstlisting}
The first column of each block contains the PDG ID \cite{ParticleDataGroup:2024cfk} of the decaying Higgs boson. The second and the third columns contain the PDG numbers of the daughter particles. The fourth columns contains the corresponding coupling $\kappa$.
The only exception to this pattern is the entry at position (PDG ID, 0, 0) in the \texttt{NORMALIZEDEFFHIGGSCOUPLINGS} block, which contains the total width of the SM-like Higgs boson of mass equal to the BSM state with a given PDG ID.
This value can be useful in the calculation of certain observables, for example to obtain the ratio of branching ratios, see the definition of $\mu_{\gamma \gamma}$ in Eq.~(10) of \cite{Kalinowski:2024uxe}, for an example.

\subsection{Treatment of invisible and undetected widths}
\label{sec:inv}

Apart from the SM-like decays described in the previous section, there is also the phenomenologically important category of invisible or undetected non-SM decays.
The former are decays to stable, neutral particles like dark matter.
The latter are other, non-SM decays that would be undetected due to large SM backgrounds.

In \FS\ potential invisible decays are identified from the set of decays into particles specified in \texttt{PotentialLSPParticles} list, defined in the model's \texttt{FlexibleSUSY.m.in} steering file.
As an example, in the MSSM one may write (see \texttt{model\_files/\allowbreak MSSM/FlexibleSUSY.m.in})
\begin{lstlisting}
PotentialLSPParticles = {Chi, Sv, Su, Sd, Se, Cha, Glu};
\end{lstlisting}
which contains all particles odd under R-parity.
If the lightest particle from this collection is electrically and color neutral, we identify a 2-body decay into a pair of those particles (or a particle and its antiparticle in case when the particle is not its own antiparticle) as an invisible decay.
Once the invisible width has been identified (or set to 0), the undetected width is computed according to
\begin{equation}
 \Gamma_\text{undet.} = \Gamma_\text{tot} - \Gamma_\text{inv.} - \sum_{i \in \SM} \Gamma_i,
\end{equation}
where the SM sum does not include potential lepton flavour violating decays.

\subsection{Interface to \HT and \LL}
\label{sec:interface}

Both \HT and \LL accept as input effective Higgs couplings to SM particles normalized to their SM values (this is sometimes referred to as reduced couplings).
These couplings are supplemented with non-SM quantities: the direct invisible and undetected decay widths (in \HT) or branching ratios (in \LL).
In case of \HT one is also allowed to specify certain additional decay widths, see below.

Based on that input, \HT and \LL provide the $\chi^2_{\BSM}$ value for the combined signal from SM-like states.
In the determination of the agreement between the parameter point and the data (quantified by the $p$-value) we follow the prescription of Ref.~\cite{Biekotter:2023oen}.
For a given program and experimental dataset that was used to compute $\chi^2_{\BSM}$, we compute the SM $\chi^2_\SM$ for the reference SM Higgs mass, which is specified in entry~1 of the \texttt{FlexibleSUSYInput} \SLHA input block (if not given, \FS uses a default value).
We assume that all reduced couplings are equal to~1 in that case.
We define $\Delta \chi^2 \coloneq \max(\chi^2_{\BSM} - \chi^2_{\SM}, 0)$, which is chi-square distributed with~2 degrees of freedom.
The $p$-value is the upper-tail probability of this distribution,
\begin{equation}
 p = 1 - \text{CDF}(\Delta \chi^2),
\end{equation}
where CDF is the cumulative distribution function.
For the case at hand $\text{CDF}(\Delta \chi^2) = 1-\exp(-\Delta \chi^2/2)$.
The~95\% CL\ corresponds therefore to $\Delta \chi^2 \approx 5.99$, whereas the $2\sigma$ exclusion corresponds to $\Delta \chi^2 \approx 6.18$.
This procedure is common to both the \HT and \LL interface.

The resulting value for $\chi^2_\BSM$, the reference value $\chi^2_\SM$ and the corresponding computed $p$-value are written to the entries~2, 3 and~4 of the \SLHA output blocks \texttt{HIGGSSIGNALS} and \texttt{LILITH}, respectively.
The output from the \WL interface from \HT has a similar form, see \secref{sec:quick}.
Entry~1 in each block contains the number of degrees of freedom from the \HT or \LL dataset.

\subsubsection{\HT}
\label{sec:ht_implementation}

As explained above, in \HT we opt to compute Higgs predictions primarily based on the effective couplings.
The loop-induced effective couplings of Higgs-gluon-gluon ($hgg$) and Higgs-photon-photon ($h\gamma\gamma$) are set according to the \FS calculation.
Note, that \HT has an option to approximate these couplings based on other effective couplings. This approach, however, yields incorrect results in the case when there are new particles contributing directly to the $h\gamma\gamma$ or $hgg$ vertices.

Higgs predictions computed based on effective couplings are supplemented with additional non-SM channels, including all quark and lepton flavours.
\FS provides flavour violating partial widths into $e^\pm \mu^\mp$, $e^\pm \tau^\mp$, $\mu^\pm \tau^\mp$, the widths for direct Higgs decay into invisible particles coming from the decay into the LSP (if present in the model) and also the reduced coupling from the 4-point interaction of a given Higgs state (the ratio of the equivalent of the SM quartic Higgs coupling parameter $\lambda$ to its SM value).

Every remaining partial width---i.e.\ widths not covered by SM decay channels, flavour violating decays or decay into the LSP---is set collectively as an \texttt{Undetected} partial width.
This tag has no particular meaning to \HT.
An \texttt{Undetected} width is only provided to signal that all branching ratios computed up to that point should be rescaled to include this extra contribution to the total width.

The theoretical predictions are calculated by \HT using \texttt{SMHiggsInterp} as the reference model.
This model interpolates between \texttt{SMHiggs} and \texttt{SMHiggsEW} depending on the provided Higgs mass (see \sectionname~4 of Ref.~\cite{Bahl:2022igd} for details) and is the current default in \HT.
In \HT the theoretical uncertainty of the mass of the decaying Higgs boson is set to~3\% of its physical mass.

\HT also checks non-SM Higgs bosons or non-SM properties of SM-like Higgs bosons (for example, the direct decay into invisible final state) via its \HB module.
The ratios to~95\% CL\ observed (excluded) cross sections is stored in entries~1 and~2, respectively, of the \SLHA output block \texttt{HIGGSBOUNDS}.

\subsubsection{\LL}
\label{sec:ll_implementation}

The input to \LL is to a large extent identical to \HT.
Similarly to \HT, we parametrize the Higgs sector in terms of effective couplings normalized to SM values.
Compared to \HT, only the~3 heaviest quark ($t$, $b$, $c$) and the~2 heaviest lepton flavours are taken into account. Flavour violating decays are not considered.
Furthermore, the Higgs self coupling is also not directly tested.
We note, that in principle another option would be to provide signal strengths. This would, however, require the calculation of cross sections, which is currently not supported by \FS.

A unique feature of \LL compared to \HT is the possibility to distinguish between production and decay for the $hgg$ or $h\gamma\gamma$ effective couplings.
We do not use this feature and set production and decay couplings for a given final/initial state to be the same, as in \HT.

The effective couplings are supplemented by branching ratios of invisible and undetected decays (not partial widths as in \HT), which are determined in the same way as in \HT.

\LL operates in two precision modes: \texttt{BEST-QCD} and \texttt{LO}, see \sectionname~4.6.2 of Ref.~\cite{Bernon:2015hsa}.
In the former mode, the imaginary part of the effective couplings is disregarded.
This is the default precision in \SA/\FS models without CP-violation in the Higgs sector.
For models with CP-violation \FS calls \LL in the \texttt{LO} mode.

Finally, instead of imposing an uncertainty on the Higgs mass $M_h$, \LL applies a hard cut on $M_h$, requiring that $M_h \in [123, 128]\GeV$.
This means that, whenever the Higgs mass lies outside of this range, the point is not checked by \LL and no exclusion is provided.
The point can still be checked by \HT, though.

\section{Example application}
\label{sec:application}

In this section, we apply \FS to the study of few popular BSM models.
First, we analyse the Higgs sector of the Type-II Two Higgs Double Model.
Then, we show how the Higgs boson properties can be used to constrain CP-violating parameters of the Next-to-Minimal Supersymmetric Standard Model.
Finally, to demonstrate the capabilities of \FS in user-defined BSM models and the treatment of invisible and undetected Higgs decays, we study the properties of the lightest Higgs boson with substantial branching ratio into a pair of stable Dirac neutralinos in the Minimal R-symmetric Supersymmetric Standard Model.

For the numerical analysis throughout this section we use \FS~2.9.0 in combination with \HT~1.2, with the \HS~1.1 and \HB~1.7 datasets.
In addition we use the development version of \LL (commit 025455fa) with the latest official dataset~19.09.
The used \LL dataset contains results from LHC Run~1, and Run~2 with $36\,\text{fb}^{-1}$, whereas \HT dataset contains also newer measurements.
The \HT and \LL datasets have 159 and 66 degrees of freedom, respectively.
For the $125.2\GeV$ SM Higgs this results in a SM $\chi^2$ of $\chi^2_{\SM,\text{\HT}} \approx 151.64$ and $\chi^2_{\SM,\text{\LL}} \approx 54.78$, respectively.
These are the values used in the determination of~95\% exclusion limits from \HS and \LL.
The procedure to obtain the exclusion limits is described in \secref{sec:interface}.

\subsection{Real Type-II 2HDM}
\label{sec:2hdm}

\FS is distributed with a ready-made Type-II 2HDM.
The scalar potential is given by
\begin{align}
    V={}& m_{11}^2\Phi_1^{\dagger}\Phi_1+m_{2}^2\Phi_2^{\dagger}\Phi_2-\Big(m_{12}^2\Phi_1^{\dagger}\Phi_2+ \text{h.c.} \Big)\nonumber\\
    &+\lambda_1\Big(\Phi_1^{\dagger}\Phi_1\Big)^2+\lambda_2\Big(\Phi_2^{\dagger}\Phi_2\Big)^2+\lambda_3\Big(\Phi_1^{\dagger}\Phi_1\Big)\Big(\Phi_2^{\dagger}\Phi_2\Big)+\lambda_4\Big(\Phi_1^{\dagger}\Phi_2\Big)\Big(\Phi_2^{\dagger}\Phi_1\Big)\nonumber\\
    &+\left[\frac{1}{2}\lambda_5\Big(\Phi_1^{\dagger}\Phi_2\Big)^2+\lambda_6\Big(\Phi_1^{\dagger}\Phi_1\Big)\Big(\Phi_1^{\dagger}\Phi_2\Big)+\lambda_7\Big(\Phi_2^{\dagger}\Phi_2\Big)\Big(\Phi_1^{\dagger}\Phi_2\Big)+ \text{h.c.}\right].
\end{align}
As the starting point of the analysis, we consider the parameter point with
\begin{gather}
\lambda_1 = -0.285, \quad
\lambda_2 = 0.128, \quad
\lambda_3 = 0.0353, \quad
\lambda_4 = -0.0506, \quad
\lambda_5 = 0.371, \quad \nonumber \\
\lambda_6 = \lambda_7 = 0, \quad
\tan \beta = 3.15, \quad
m_{12} = 250\GeV,
\label{eq:2hdm_bmp}
\end{gather}
where the parameters are defined in the $\overline{\abbrev{MS}}$ scheme at the scale $Q = 125.2\GeV$.
Based on a full 1-loop calculation, \FS predicts the following physical masses
\begin{equation}
m_h = 125\GeV, \quad m_H = 460\GeV, \quad m_A = 440\GeV, \quad m_{H^+} = 455\GeV.
\end{equation}
This parameter point has $\chi^2 = 151.54$ in \HT and $\chi^2=54.70$ in \LL.
Both of these values are smaller than the corresponding SM $\chi^2$, see above, which means that \FS reports $p$-values of $p=1$.
The input and the output spectrum files for this point are attached to the \texttt{arXiv} version of this work.

\begin{figure}
  \centering
  \includegraphics[width=0.9\textwidth]{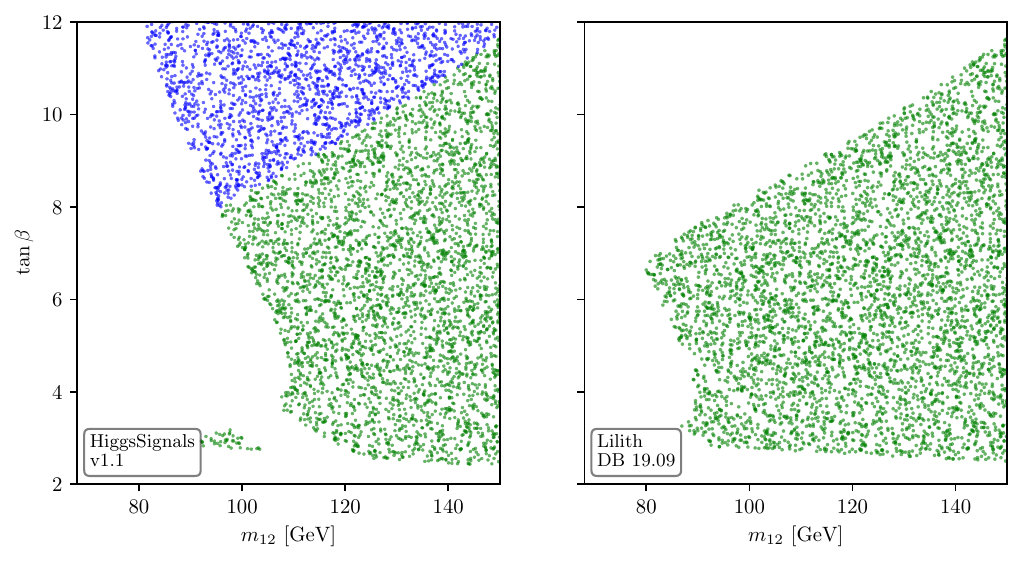}
  \caption{Scan over $m_{12}$ and $\tan \beta$ in the vicinity of the
    2HDM point defined in Eqs.~\eqref{eq:2hdm_bmp}. All shown points
    are allowed by \HS (left) or \LL (right) at~95\% CL. Blue points
    have a Higgs mass outside of the $[123,128]\GeV$ interval, but are
    nevertheless allowed by \HT with~3\% mass uncertainty.}
  \label{fig:thdm_hs}
\end{figure}

To apply the constraining power of Higgs data, we scan over the parameters $m_{12}$ and $\tan \beta$ around the point specified in Eqs.~\eqref{eq:2hdm_bmp}.
In \figref{fig:thdm_hs} we zoom in on the region where $m_{12} \in [80, 150]\GeV$.
All shown points are consistent with the experimental data at~95\% CL according to \HT (left plot) and \LL (right plot).
The origin of the difference between the left and the right plot is two fold.
First, the uncertainty of the Higgs mass is treated differently in both programs.
In the \HT interface we specify a~3\% uncertainty, see \sectionname~2.2.3 of Ref.~\cite{Bahl:2022igd} for details.
Such a possibility does not exist in \LL.
As described in the previous section, \LL applies a hard cut of $M_h$ to the $[123,128]\GeV$ range.
This distinction is illustrated in the left plot of \figref{fig:thdm_hs}, where the blue points are allowed according to \HT but have a SM-like Higgs mass which falls outside of the range that would be accepted by \LL.
Even all other settings being equal, these points would be excluded by \LL by the Higgs mass cut.
Other then that, the difference comes from different experimental analysis implemented in \HT and \LL, where \HT includes newer data and is therefore more constraining.
This is especially visible in the bottom-left part of the left plot.

\begin{figure}
  \centering
  \includegraphics[width=0.9\textwidth]{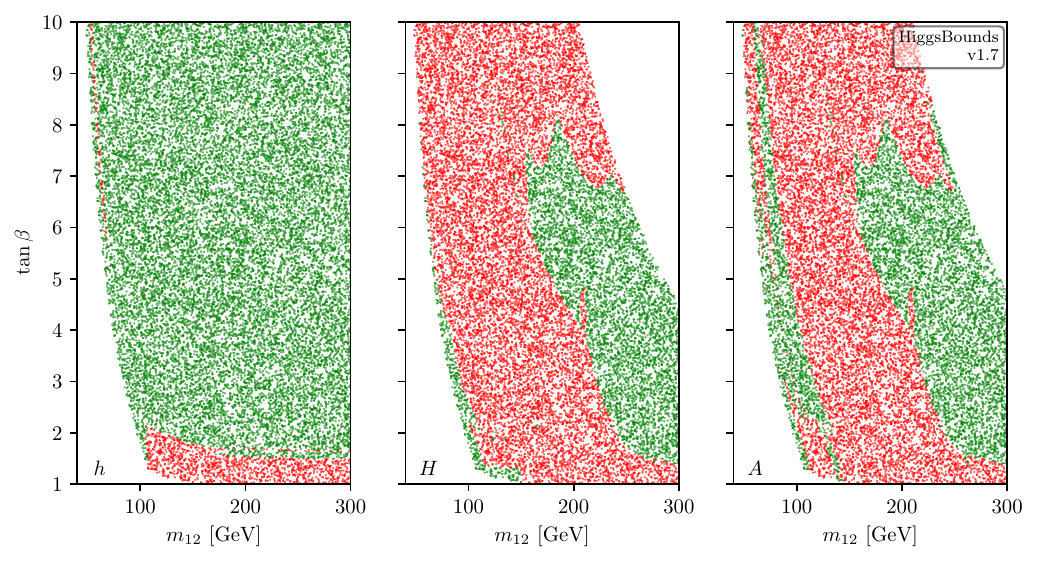}
  \caption{Scan over $m_{12}$ and $\tan \beta$, as in
    \figref{fig:thdm_hs}. Green (red) points are allowed (excluded) by
    \HB at~95\% CL by the production of the lightest CP-even Higgs
    $h$ (left), the next-to-lightest CP-even Higgs $H$ (middle) and
    the pseudoscalar Higgs $A$ (right).}
  \label{fig:thdm_hb}
\end{figure}

In \figref{fig:thdm_hb} we focus on non-SM properties of $h$ and searches for non-SM Higgs bosons $H$ and $A$.
This is analysed by the \HB module of \HT.
\HB reports~95\% CL agreement with observed and expected limits for individual BSM states (see \secref{sec:quick}), in our case: $h$, $H$ and $A$.
Red points in \figref{fig:thdm_hb} are excluded at~95\% CL (observed) by properties of $h$ (left), $H$ (middle) and $A$ (right panel).
We give few example cases of why they are excluded.
Starting with the left plot, red points correspond to cases where $h$ becomes non-SM like.
In the bottom part of the left plot $h$ becomes so light that it is excluded, for example, by searches at LEP \cite{LEPWorkingGroupforHiggsbosonsearches:2003ing} or by searches for low mass Higgses at the LHC \cite{CMS:2024yhz}.
On the other hand, the heavy Higgses from the middle and right panel, are mostly constrained by searches for heavy resonances at the LHC.
For example, by searches in $ZZ$ channel \cite{ATLAS:2020tlo} for heavy Higgs $H$ or in $\tau^+ \tau^-$ \cite{CMS:2022goy} for $A$.

We note that, when using \HT in real life analysis, the allowed parameter space is the overlap of the allowed regions from \figref{fig:thdm_hs} and~\ref{fig:thdm_hb}.
The decision of how to combine them is left to the user, which is provided with all the needed information in the output.

\subsection{CP-violating NMSSM}
\label{sec:NMSSMCPV}

The SM predicts that the Higgs boson is a CP-even scalar.
Despite that, many BSM models predict also the existence of CP-odd Higgs-like states.
The hypothesis, that the Higgs boson with $M_h\approx 125\GeV$ observed at the LHC is such a pure CP-odd scalar is currently disfavoured at the level of $3.4\sigma$ by ATLAS \cite{ATLAS:2022akr} and $3.0\sigma$ by CMS \cite{CMS:2021sdq}.
However, it may still have a substantial admixture of a CP-odd component.

Experimentally the CP properties of a Higgs can be accessed through the analyses of angular and kinematic correlations in its production and decay, which are sensitive to the tensor structure of Higgs interactions.
The strongest constraint in this respect comes from the measurement of the $h\to\tau\bar{\tau}$ decay.
The general coupling of $\tau$ to Higgs has has the form
\begin{equation}
 \mathcal{L} \ni -\frac{m_\tau}{v} \kappa_\tau \bar{\tau} \left[\cos(\phi_\tau) + \imath\sin(\phi_\tau)\gamma^5\right]  \tau h,
\end{equation}
where $\kappa_\tau$ is the reduced Yukawa coupling strength and $\phi_\tau \in [-\pi/2,\pi/2]$ is the CP mixing angle that parametrises the relative contributions of the CP-even and CP-odd components to the $h\tau\tau$ coupling.
The SM CP-even state corresponds to $\kappa_\tau =1$ and $\phi_\tau = 0$, while a pure CP-odd state would imply $\phi_\tau = \pm\pi/2$.
A recent measurement determined this angle to be $(9 \pm 16)^\circ$ \cite{ATLAS:2022akr}, which is consistent with the SM but still leaves room for a substantial CP-odd admixture.
The similar determination made for the coupling to the top quark from the $t\bar{t}h$ production yields $\phi_t = 11^{\circ+52^\circ}_{~-73^\circ}$ \cite{ATLAS:2023cbt}.

Currently, to our knowledge, the CMS analysis of $h\to\tau^+\tau^-$ \cite{CMS:2021sdq} is the only analysis in \HS targeting CP-odd structure of Higgs-fermion coupling directly, based on a dedicated CP-odd observable.
No similar analysis is currently present in \LL.

To show how Higgs measurements can give access to underlying CP-violating parameters of the BSM model we focus on a CP-violating scenario which can enhance this particular observable.
This is possible thanks to the fact that \FS fully supports complex parameters allowing for the study of many interesting models with new sources of CP violation since version 2.0.0 \cite{Athron:2017fvs}.
To showcase the constraining power of the Higgs data we consider the high-scale, CP-violating, $Z_3$-symmetric NMSSM.
This is one the simplest supersymmetric models which features CP-violation in the Higgs sector already at the tree level.

We take the NMSSM superpotential to be
\begin{equation}
	W_\NMSSM = W_{\MSSM}(\mu=0) + \lambda \hat{S}\hat{H}_u\cdot\hat{H}_d + \frac{\kappa}{3} \hat{S}^3,
\end{equation}
where $W_{\MSSM}(\mu=0)$ is the MSSM superpotential with the $\mu$-parameter set to zero.
The phases of scalar Higgs fields, responsible for CP-violation in tree-level potential, which are the most relevant parameters for this analysis, are defined as
\begin{align}
  H_u &= \ee^{\imath \eta} \colvec{H_u^+ \\ (v_u + \phi_u + \imath \sigma_u)/\sqrt{2}}, &
  H_d &= \colvec{(v_d + \phi_d + \imath \sigma_d)/\sqrt{2} \\ H_d^-}, &
  S &= \frac{\ee^{\imath \eta_S}}{\sqrt{2}} (v_S + \phi_S + \imath \sigma_S).
  \label{eq:doubletDefs}
\end{align}
We consider the NMSSM parameter point given by
\begin{gather}
   m_0 = 940\GeV, \quad m_{1/2} = 851\GeV, \quad A_0 = -2666\GeV, \quad \tan \beta = 11, \nonumber \\
   \lambda = -0.32,~\quad \kappa = 0.64,~\quad A_\lambda = -370\GeV, \quad A_\kappa = 31\GeV \quad \lambda v_S = -161\GeV, \nonumber \\
   \eta = -1, \quad \eta_s = -2.77.
  \label{eq:NMSSMBMP}
\end{gather}
At 1-loop level \FS predicts the Higgs masses
\begin{gather}
  m_{H_1} = 63\GeV, \quad m_{H_2} = 125.2\GeV, \quad m_{H_3} = 345\GeV, \quad m_{H_3} = 858\GeV, \nonumber \\
  m_{H_4} = 859\GeV, \quad m_{H^+} = 860\GeV,
\end{gather}
which yields $\chi^2 = 156.56$ from \HS with 159 degrees of freedom.
We note that this result contains the analysis from Ref.~\cite{CMS:2021sdq}, which provides 3 degrees of freedom. Its removal would reduce $\chi^2$ to $153.92$.
The reference SM $\chi^2$ without \cite{CMS:2021sdq} is $\chi^2_{\SM}=149.30$.
This corresponds to a $p$-value of $p=0.085$ with or $p=0.099$ without analysis from Ref.~\cite{CMS:2021sdq}.
The input and output spectrum files for this point, together with \HS outputs in both configurations, are attached to the \texttt{arXiv} version of this publication.

\begin{figure}
 \centering
 \includegraphics[width=0.6\textwidth]{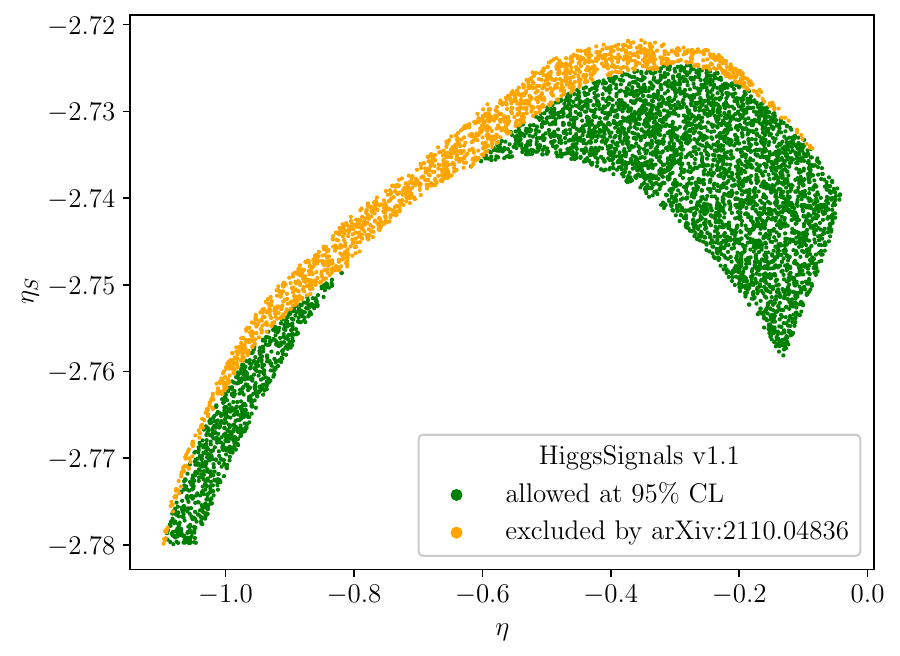}
 \caption{Scan over NMSSM CP-violating phases $\eta$ and $\eta_S$
   around the parameter point given in Eq.~\eqref{eq:NMSSMBMP}.  Green
   points are allowed by \HS at~95\% CL. Yellow points are allowed
   if the CMS analysis of the CP structure of the Yukawa coupling
   between the Higgs boson and $\tau$ leptons \cite{CMS:2021sdq} is
   excluded.}
 \label{fig:cp}
\end{figure}

To show the usefulness of CP-sensitive observables, we show in \figref{fig:cp} a scan over the CP-violating phases $\eta$ and $\eta_S$ from Eqs.~\eqref{eq:doubletDefs} around the parameter point given in Eqs.~\eqref{eq:NMSSMBMP}.
The green points are allowed by \HS at~95\% CL.
The yellow points are excluded when including the constraints from Ref.~\cite{CMS:2021sdq}, but are allowed otherwise.
This scan is directly sensitive to the CP properties of the NMSSM Higgs sector.
It avoids the problem, that it is impossible to determine the CP status of the Higgs just from production and decay rates.
This issue can be seen by considering a fermionic decay channel, as given in Eq.~\eqref{eq:fermion_partial}.
The same partial width can be obtained in the CP-conserving case by the replacement $\tilde{c}_f\to 0$ and
afterwards $c_f^2 \to c_f^2 + \tilde{c}_f^2/\beta^2$.
Without looking at CP-sensitive observables it is therefore impossible to say whether the deviation comes from a CP-violating effect or from altered values of CP-even couplings.

\subsection{Invisible Higgs decay in the MRSSM}

Numerous BSM models, especially those providing dark matter candidates, can feature a substantial invisible Higgs decay width.
Such is the case, for example, in popular models like the Inert 2HDM or the MSSM.
This final state is currently relative weakly constrained by experimental measurement, with the observed upper limit at the~95\% CL for $\BR(H \to \text{inv})$ of the $125\GeV$ Higgs set at 0.107 by ATLAS \cite{ATLAS:2023tkt} and 0.15 by CMS \cite{CMS:2023sdw}, leaving substantial room for a potential BSM contribution.

One of the models that can provide this signature is the Minimal R-symmetric Supersymmetric Standard Model (MRSSM) \cite{Kribs:2007ac}.
The model features a Dirac neutralino $\chi$ which might be light, especially in scenarios where the SM-like Higgs is accompanied by a second mostly singlet like state of lower mass, as explored in Refs.~\cite{Diessner:2015iln,Kalinowski:2024uxe}.
It is a very interesting model for which no dedicated tools exists, making it a perfect candidate to showcase the ability to study Higgs constraints in user defined BSM models in \FS.

\begin{table}
\begin{center}
\begin{tabular}{c|l|l||l|l|l|l}
\multicolumn{1}{c|}{Field} & \multicolumn{2}{c||}{Superfield} &
                              \multicolumn{2}{c|}{Boson} &
                              \multicolumn{2}{c}{Fermion} \\
\hline
 \phantom{\rule{0cm}{5mm}}Gauge Vector    &\, $\hat{g},\hat{W},\hat{B}$        \,& \, $\;\,$ 0 \,
          &\, $g,W,B$                 \,& \, $\;\,$ 0 \,
          &\, $\tilde{g},\tilde{W}\tilde{B}$             \,& \, +1 \,  \\
Matter   &\, $\hat{l}, \hat{e}$                    \,& \,\;+1 \,
          &\, $\tilde{l},\tilde{e}^*_R$                 \,& \, +1 \,
          &\, $l,e^*_R$                                 \,& $\;\;\,$\,\;0 \,    \\
          &\, $\hat{q},{\hat{d}},{\hat{u}}$       \,& \,\;+1 \,
          &\, $\tilde{q},{\tilde{d}}^*_R,{\tilde{u}}^*_R$ \,& \, +1 \,
          &\, $q,d^*_R,u^*_R$                             \,& $\;\;\,$\,\;0 \,    \\
 $H$-Higgs    &\, ${\hat{H}}_{d,u}$   \,& $\;\;\,$\, 0 \,
          &\, $H_{d,u}$               \,& $\;\;\,$\, 0 \,
          &\, ${\tilde{H}}_{d,u}$     \,& \, $-$1 \, \\ \hline
\phantom{\rule{0cm}{5mm}} R-Higgs    &\, ${\hat{R}}_{d,u}$   \,& \, +2 \,
          &\, $R_{d,u}$               \,& \, +2 \,
          &\, ${\tilde{R}}_{d,u}$     \,& \, +1 \, \\
  Adjoint Chiral  &\, $\hat{O},\hat{T},\hat{S}$     \,& \, $\;\,$ 0 \,
          &\, $O,T,S$                \,& \, $\;\,$ 0 \,
          &\, $\tilde{O},\tilde{T},\tilde{S}$          \,& \, $-$1 \,  \\
\hline
\end{tabular}
\end{center}
\caption{The R-charges of the superfields and the corresponding
  bosonic and fermionic components.}
\label{tab:Rcharges}
\end{table}
The main principle behind the MRSSM is that it is a supersymmetrized 2HDM with an unbroken R-symmetry \cite{Fayet:1974pd}.
Original interest in this model stem from its flavour safety even in the presence of large, off-diagonal soft SUSY-breaking mass terms \cite{Kribs:2007ac}, as R-symmetry forbids the MSSM $A$-terms.
Importantly, soft Majorana gaugino mass terms are also forbidden by R-symmetry.
However, the fermionic components of the chiral adjoints, $\hat{\Phi}_i\in\{\hat{O},\hat{T},\hat{S}\}$
for each standard model gauge group $i\in\{SU(3),SU(2),U(1)\}$ respectively, can be paired with standard gauginos $\tilde{g}$, $\tilde{W}$, $\tilde{B}$ to build Dirac fermions and the corresponding mass terms.
The same issue occurs for the $\mu$ term, whose counterpart can be created thanks to the addition of the so-called R-Higgses, which carry R-charge of~2.
The field content of the model, together with R-charge assignments, is summarized in \tabref{tab:Rcharges}, where superfields not present in the MSSM are shown in the bottom part of the table.

The MRSSM superpotential is given by
\begin{align}
  W ={}& \mu_d\,\hat{R}_d \cdot \hat{H}_d\,+\mu_u\,\hat{R}_u\cdot\hat{H}_u\,+\Lambda_d\,\hat{R}_d\cdot \hat{T}\,\hat{H}_d\,+\Lambda_u\,\hat{R}_u\cdot\hat{T}\,\hat{H}_u \nonumber \\
       & +\lambda_d\,\hat{S}\,\hat{R}_d\cdot\hat{H}_d\,+\lambda_u\,\hat{S}\,\hat{R}_u\cdot\hat{H}_u\,
         - Y_d \,\hat{D}\,\hat{Q}\cdot\hat{H}_d\,- Y_e \,\hat{E}\,\hat{L}\cdot\hat{H}_d\, +Y_u\,\hat{U}\,\hat{Q}\cdot\hat{H}_u\, .
\label{eq:superpot}
\end{align}
It consists of MSSM Yukawa terms, substitutes for the MSSM $\mu$-term, which is forbidden, and new Yukawa-like interactions with $U(1)$ and $SU(2)$ adjoints.
Of importance to this analysis are mostly the terms proportional to $\mu_u$ and $\lambda_u$, as explained below.
We refer to Ref.~\cite{Diessner:2014ksa} for a detailed description of the model and the discussion of remaining parameters, and especially their influence on the Higgs mass and precision EW observables.

The model provides four Dirac neutralinos $\chi_i$, the lightest of which $\chi_1$ can be stable thanks to R-symmetry, making it a dark matter candidate.
The neutralino mass matrix, in the basis
$ {\xi}_i=({\tilde{B}}, \tilde{W}^0, \tilde{R}_d^0, \tilde{R}_u^0)$,  $\zeta_i=(\tilde{S}, \tilde{T}^0, \tilde{H}_d^0, \tilde{H}_u^0)$, has the following form
\newcommand{\muu}[1]{\mu_u^{\text{eff,}#1}}
\newcommand{\mud}[1]{\mu_d^{\text{eff,}#1}}
\begin{equation}
\label{eq:neut-massmatrix}
m_{{\chi}} = 
\begin{pmatrix}
M^{D}_B &0 &-\frac{1}{2} g_1 v_d  &\frac{1}{2} g_1 v_u \\
0 &M^{D}_W &\frac{1}{2} g_2 v_d  &-\frac{1}{2} g_2 v_u \\
- \frac{1}{\sqrt{2}} \lambda_d v_d  &-\frac{1}{2} \Lambda_d v_d  & - \mud{+} &0\\
\frac{1}{\sqrt{2}} \lambda_u v_u  &-\frac{1}{2} \Lambda_u v_u  &0 & \muu{-}
\end{pmatrix},
\end{equation}
where
\begin{equation}
\mu_i^{\text{eff,}\pm} =\mu_i+\frac{\lambda_iv_S}{\sqrt2}
\pm\frac{\Lambda_iv_T}{2}
\end{equation}
and where $v_S$ and $v_T$ are the vacuum expectation values (VEVs) of the CP-even parts of the scalar components of electroweak gauge adjoints.
The transformation to a diagonal mass matrix and mass eigenstates $\kappa_i$ and $\psi_i$
is performed by two unitary mixing matrices \(N^1\) and \(N^2\) as
\begin{equation}
  N^{1*} m_{{\chi}} N^{2\dagger} = m^\text{diag}_{{\chi}}, \qquad
  {\xi}_i=\sum_{j=1}^4 N^{1*}_{ji}{\kappa}_j, \qquad
  \zeta_i=\sum_{j=1}^4 N^{2*}_{ij}{\psi}_j,
\end{equation}
and physical four-component Dirac neutralinos are constructed as
\begin{equation}
  \chi^0_i= \colvec{\kappa_i\\ {\psi}^{*}_i}, \qquad i\in\{1,2,3,4\}.
  \label{eq:neat-mix}
\end{equation}
In the case when the Dirac mass parameter $M^D_B$ is small compared to other dimensionful parameters, the lightest neutralino is a mostly a Bino-Singlino state.
Its phenomenology as a dark matter candidate has been extensively studied in Ref.~\cite{Diessner:2015iln}, and re-analysed using new dark matter direct detection constraints in the context of a potential $95\GeV$ scalar excess in Ref.~\cite{Kalinowski:2024uxe}.
Here we will only briefly remind details relevant to the analysis at hand.

In the limit where $\mu_d^{\text{eff},+}, M^D_W \gg M^D_B, \mu_d^{\text{eff},+}, v_u, v_d$ the fields $\xi_2$, $\xi_3$, $\zeta_2$ and $\zeta_3$ decouple and the mixing of the remaining fields is described by the mass matrix
\begin{equation}
m_\chi \approx
\begin{pmatrix}
M^D_{B} & \frac{1}{2} g_1 v_u \\
\frac{1}{\sqrt{2}} \lambda_u v_u & \mu_u + \frac{1}{\sqrt{2}}\lambda_u v_S
\end{pmatrix},
\end{equation}
where we have also neglected terms proportional to $v_T$ as this VEV is constrained to be the electroweak precision data to be $\mathcal{O}(1\GeV)$.
The mass of the lightest state, which is mostly a Bino-Singlino with a small admixture of an up-type (R-)Higgsino, is in this approximation given by
\begin{equation}
 m_{\chi_1} \approx M_B^D - \frac{g_1 \lambda_u v_u^2}{2 \sqrt{2}\mu_u}\,.
\end{equation}
For the parameter space we consider, this correction is small so at the tree-level with a good approximation the mass of the dark matter particle is given by the value of $M^D_B$.

As usual, the SM-like Higgs $H_1$ is composed primarily of $H_u$.
The coupling between $H_1$ and $\chi_1$ can be approximated by
\begin{equation}
 H_1 \bar{\chi}_1 \chi \approx - Z_{12}^H \left[\left(\sqrt{2} \lambda_u N^{1}_{14}  N^{2}_{11} - g_1 N^{1*}_{11} N^{2*}_{14} \right) P_L + \left(\sqrt{2} \lambda_u N^1_{14}  N^2_{11} - g_1 N^1_{11} N^2_{14} \right) P_R\right],
\end{equation}
where we have omitted potential mixing with heavy components.
Physically, it means that a Bino-Singlino state couples to $H_u$ either via the Bino component through gauge-kinetic term (the terms $\propto g_1 N^{1}_{11} N^{2}_{14}$) or via the Singlino component through a $\lambda_u$ interaction term in the superpotential given in Eq.~\eqref{eq:superpot} (the terms $\propto \lambda_u N^{1}_{14} N^{2}_{11}$).


To show some phenomenological consequences of this setup, we consider the parameter point defined by
\begin{gather}
\lambda_u = 0.1, \quad \lambda_d = -1.4, \quad \Lambda_u = -1.3,  \quad \Lambda_d = 1.1, \nonumber \\
M_B^D = 30\GeV, \quad \mu_u = 250\GeV,  \quad \tan \beta = 14, \quad B_\mu = (670\GeV)^2, \nonumber \\
m_q = 1.75\TeV, \quad m_l = m_e = 1\TeV, \quad m_{R_u} = 500\GeV, \quad m_S = 400\GeV, \nonumber \\
\mu_d = M_W^D = M_O^D = m_T = 3\TeV, \quad m_O = 1.5\TeV, \quad m_{R_d} = 1500\GeV.
\label{eq:MRSSMBMP}
\end{gather}
Soft-mass parameters which are not relevant to this analysis, like the gluino mass parameter $M_O^D$, are set to values at or well above $1\TeV$.
This mostly removes those degrees of freedom from consideration.
This parameter point gives $m_{H_1} = 125.16\GeV$ and $m_{\chi_1} = 23.9\GeV$, with a~42\% branching ratio of $H_1$ to $\bar{\chi}_1 \chi_1$.
The only other light Higgs state is $A_1$, with mass of $185\GeV$.
The remaining Higgses have masses above $800\GeV$. The \SLHA\ input and \FS outputs for this point, in the two configurations with and without taking into account invisible decay width, are attached to the \texttt{arXiv} version of this publication.

As explained in \secref{sec:inv}, the invisible decay width of a given Higgs is identified in \FS as the decay to lightest, odd, charge and colour neutral particles, if such exists in the model.
In the scenario considered, \FS automatically identifies it as $H_1 \to \bar{\chi}_1 \chi_1$ for the decay of the lightest Higgs boson.
At the same time, the undetected branching ratio of $H_1$ is equal to~0 for this point.

Despite this, the normalized effective couplings $\kappa$ are SM-like.
Rounded to~2 decimal places, we obtain
\begin{align}
  \kappa_Z &= 1, & \kappa_W &= 1, & \kappa_g &= 1, & \kappa_\gamma &= 1.06, & \kappa_{\gamma Z} &= 1.02, & \kappa_t &= 1, & \kappa_\tau &= 0.97,
  \label{eq:effCvals}
\end{align}
which would make the point experimentally allowed if not for the invisible decay mode of $H_1$.
Its inclusion systematically lowers all SM branching ratios of $H_1$, with the BR to a concrete final state $f \in \{\gamma \gamma, \tau^+ \tau^-, ZZ, \ldots\}$ rescaled in the presence of an invisible partial width $\Gamma_\text{inv}$ or/and a hypothetical undetected partial width $\Gamma_\text{und}$ as
\begin{equation}
\BR(H_i \to f) = \frac{\Gamma(H_i \to f)}{\Gamma_\text{SM} + \Gamma_\text{inv} + \Gamma_\text{und}} \approx (1 - \BR_\text{inv} - \BR_\text{und}) \BR_\text{SM}(H_i \to f)\,,
\label{eq:scaling}
\end{equation}
where $\Gamma_\text{SM}$ represents combined decay width into SM final states and $\BR_\text{SM}$ a branching ratio calculated assuming only SM decays contribute to the total decay width.
That is why this parameter point has $p$-values $p_\text{\HS} \approx  p_\text{\LL} \approx 0$.
Artificially setting that invisible width to zero would give $p_\text{\HS} \approx 0.56$ and $p_\text{\LL} = 1$, making the point consistent with current experimental measurements.
This shows the phenomenological importance of the correct treatment of such decay modes.

\begin{figure}
 \centering
 \includegraphics[width=0.9\textwidth]{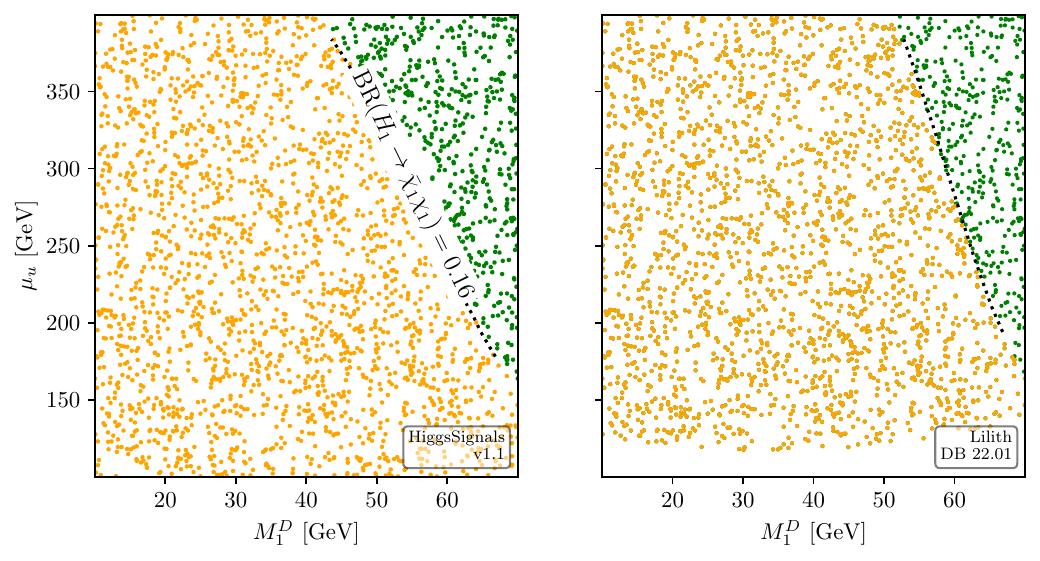}
 \caption{Parameter region excluded by the inclusion of invisible
   Higgs decay width as determined by \HS (left) and \LL (right).
   Green point are always allowed, yellow points are allowed only when
   setting the $\BR(H_1 \to \text{inv}) = 0$ and
   $\BR(H_1 \to \text{und}) = 0$ (there are no undetected decay in
   this scenario other than in the bottom-left corners).  The bottom
   half of the right plot is empty since in this region the Higgs mass
   falls outside of the hard cut region $[123,128]\GeV$ applied by
   \LL.}
 \label{fig:mrssm_hs}
\end{figure}

In \figref{fig:mrssm_hs} we study this in detail, by showing the impact of invisible Higgs decay on the constraints for the SM-like $125\GeV$ Higgs predicted by the MRSSM.
We scan over the Bino-Singlino mass parameter $M_B^D$ and the up (R-)Higgino parameter $\mu_u$ around the point specified in Eqs.~\eqref{eq:MRSSMBMP}.
These two parameters control the mass of the Bino-Singlino neutralino and its coupling to the SM-like Higgs boson.
Smaller values of $\mu_u$ and $M_B^D$ increase the $\Gamma(H_1 \to \bar{\chi}_1 \chi)$.
Yellow points in \figref{fig:mrssm_hs} are excluded at~95\% CL by the inclusions of invisible decay widths, but would be allowed otherwise while keeping all the effective couplings unchanged.

The difference in lower portions of the plots is due to the fact that \LL does not constraint points with $\mu_u \lesssim 300\GeV$, for which the Higgs mass falls outside of hard-cut region $[123,128]\GeV$ implemented by \LL.

Points in the bottom-left corners have also an undetected decay width.
For very low $\mu_u$, up-type (R-)Higgsino neutralino drops in mass below the Higgs mass, making the decay to it and a very light Bino-Singlino neutralino kinematically allowed.
This fact does not influence the shape of boundary between regions.

We note that the invisible decay width is currently very weakly constrained by direct experimental measurements as the total decay width is not known very precisely, determined only to be $\Gamma_H = 4.4^{3.0}_{-2.2}\MeV$ by ATLAS \cite{ATLAS:2023dnm} and $3.9^{2.7}_{-2.2}\MeV$ by CMS \cite{CMS-PAS-HIG-24-011}.
The impact of an inclusion of invisible and undetected partial widths on the exclusions reported by \HS and \LL comes therefore mostly from the rescaling effect shown in Eq.~\eqref{eq:scaling}.
Future $e^+e^-$ colliders, which are able to measure the invisible Higgs decay directly via the recoil-mass, are projected to narrow it down, in conjunction with HL-LHC data, to~0.2--0.3\% at~95\% CL\ \cite{deBlas:2019rxi}.
This would then allow to probe invisible partial width directly.
The current relatively low sensitivity to invisible width can be seen from the fact that the excluded points have an invisible branching ratio of more then~16\% (\HS) and~10\% (\LL), as marked by the black contour.

\begin{figure}
  \centering
 \includegraphics[width=0.6\textwidth]{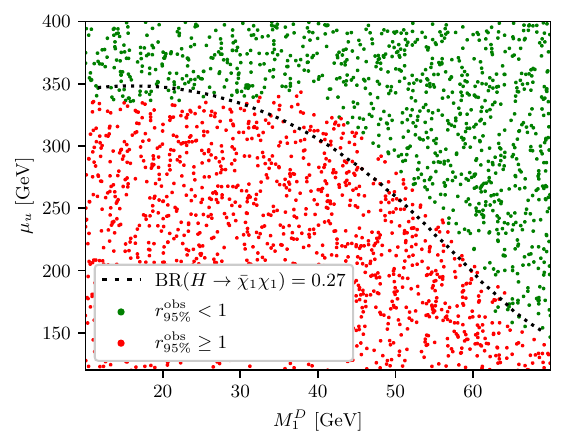}
 \caption{Study of the \HB exclusion around the point from
   Eq.~\eqref{eq:MRSSMBMP}.  Green/red points show
   $r_{95\%}^\text{obs}$, the model-predicted rate divided by the
   observed~95\% CL. Bottom-left region of the plot is exclude by
   the combined ATLAS searches for invisible decays
   \cite{ATLAS:2023tkt}.  The dashed line shows the
   $\BR(H_1 \to \bar{\chi}_1 \chi_1) = 0.27$.}
 \label{fig:hb_directInv}
\end{figure}

We explore this more in \figref{fig:hb_directInv}, where we focus solely on the direct search of invisible Higgs decays.
The setup for the scan is the same as in  \figref{fig:mrssm_hs}.
Green (red) points are allowed (excluded) by \HB at~95\% CL.
The bottom-left region of the plot is excluded mostly by the ATLAS combination of searches of direct invisible decays of a Higgs boson \cite{ATLAS:2023tkt}.
The border between these regions corresponds roughly to $\BR(H_1 \to \bar{\chi}_1 \chi_1) \approx 0.27$.
This corresponds roughly to $\Gamma(H_1 \to \bar{\chi}_1 \chi_1) \approx 1.5\MeV$, since the total Higgs width is relatively constant  on the boundary (the variation over the considered $M^D_B$ range is below~15\%).
The shape of the boundary roughly follows the expected leading $1/\mu_u$ behaviour of the partial width
\begin{equation}
  \Gamma(H_1 \to \bar{\chi}_1 \chi_1) \sim \left[1 - \left(\frac{2 M^D_B}{m_{H_1}}\right)^2\right]^{3/2} \left( \frac{g_1 \lambda_u v_u}{\mu_u} Z^H_{12}\right)^2.
\end{equation}
The quoted BR limit of~0.27 broadly coincides with the \HB limit~0.228, reported without the use \FS for a Higgs of mass $125.16\GeV$ with~3\% mass uncertainty and all normalized effective couplings set to~1.
The difference could be explained by the fact that for actual parameter points the effective couplings are not exactly~1.


\section{Conclusions}
\label{sec:summary}

Current measurements of the $125\GeV$ Higgs boson properties and searches for non SM-like BSM Higgs bosons put very strong constraints on any extension of the SM model, with incoming data from existing and planned experiments having the potential to further limit the already constrained parameter space.
To exploit this progress, we present here the new feature of the \FS spectrum-generator generator --- the interface to computer programs \HT and \LL, which confronts predicted properties of BSM Higgs boson with experimental data.
Combined with \FS's capability to evaluate Higgs boson decays with high degree of accuracy, this allows to determine the validity of the Higgs sectors in a user defined BSM models directly from within \FS.
The procedure is fully automated and requires only a minimal initial setup from the user.
Users can therefore create their own BSM models, starting from a definition of its field content and a Lagrangian, and in a fully automated way check their agreement with available Higgs data.

The input to \HT and \LL is provided by \FS in the form of effective couplings normalized to their respective SM values.
Those normalized couplings are printed out to the terminal in the \SLHA format or to a \WL\ interface via the \texttt{LibraryLink} interface of \FS.
They then can be manual inspected by the user or used in various phenomenological analysis like, for example, in the rescaling of SM production cross-sections to BSM results.

We showcased the application of the interface on three representative examples of BSM physics to illustrate the capabilities and universality of \FS.
We considered the Type-II 2HDM, the CP-violating NMSSM and the Minimal R-symmetric Supersymmetric Standard Model.
This shows that \FS can handle both SUSY and non-SUSY models as well as correctly treat models with CP-violating Higgs sectors and invisible Higgs decay widths.

While \LL checks only neutral Higgs bosons, \HT can provide constraints also for the single and double charged states.
Currently, however, the \FS interface to \HT only analyses neutral Higgs bosons.
Moreover, since we need to construct a mock-up SM to normalized BSM effective couplings, the interface checks only particles with masses $\lesssim 650\GeV$, beyond which the SM quartic Higgs coupling $\lambda$ in the model used for normalization becomes non-perturbative.
Some of these limitations could be removed by providing \HT also with explicit values of cross sections and decay widths, without the need to normalize them to the SM.
This is an alternative input accepted by \HT.
Since currently \FS does not compute cross sections, these issues will be addressed in later updates.

Further improvements to the precision of Higgs decay calculation in \FS are under way.
This includes genuine 1-loop BSM corrections to the tree-level decays in the decoupling renormalization scheme \cite{Lang:2025eom} as well as the 2-loop corrections to Higgs bosons masses based on \cite{Goodsell:2015ira}.
These refinements will allow to maximally capitalize on the \FS extension presented in this work as well as current and upcoming experimental data.

\section*{Acknowledgements}

We thank Henning Bahl and Sven Heinemeyer for their help regarding \HT, Sabine Kraml for her help with \LL and Peter Athron for his insight into the proper statistical interpretation of \HS results.

WK was supported by the National Science Centre (Poland) grant 2022/\allowbreak47/\allowbreak D/\allowbreak ST2/\allowbreak03087.

\appendix

\printbibliography[title={References}]

@article{Cepeda:2019klc,
    author = "Cepeda, M. and others",
    editor = "Dainese, Andrea and Mangano, Michelangelo and Meyer, Andreas B. and Nisati, Aleandro and Salam, Gavin and Vesterinen, Mika Anton",
    title = "{Report from Working Group 2}: {Higgs Physics at the HL-LHC and HE-LHC}",
    eprint = "1902.00134",
    archivePrefix = "arXiv",
    primaryClass = "hep-ph",
    reportNumber = "CERN-LPCC-2018-04",
    doi = "10.23731/CYRM-2019-007.221",
    journal = "CERN Yellow Rep. Monogr.",
    volume = "7",
    pages = "221--584",
    year = "2019"
}

@article{FCC:2018byv,
    author = "Abada, A. and others",
    collaboration = "FCC",
    title = "{FCC Physics Opportunities}: {Future Circular Collider Conceptual Design Report Volume 1}",
    reportNumber = "CERN-ACC-2018-0056",
    doi = "10.1140/epjc/s10052-019-6904-3",
    journal = "Eur. Phys. J. C",
    volume = "79",
    number = "6",
    pages = "474",
    year = "2019"
}

@article{Bambade:2019fyw,
    author = "Bambade, Philip and others",
    title = "{The International Linear Collider: A Global Project}",
    eprint = "1903.01629",
    archivePrefix = "arXiv",
    primaryClass = "hep-ex",
    reportNumber = "DESY 19-037, DESY-19-037, FERMILAB-FN-1067-PPD, IFIC/19-10, IRFU-19-10,
  JLAB-PHY-19-2854, KEK Preprint 2018-92, JLAB-PHY-19-2854, KEK
  Preprint 2018-92, LAL/RT 19-001, PNNL-SA-142168,
  SLAC-PUB-17412, SLAC-PUB-17412",
    month = "3",
    year = "2019"
}

@book{EuropeanStrategyGroup:2020pow,
    collaboration = "European Strategy~Group",
    title = "{2020 Update of the European Strategy for Particle Physics}",
    reportNumber = "CERN-ESU-013, CERN-ESU-015",
    doi = "10.17181/ESU2020",
    isbn = "978-92-9083-575-2",
    publisher = "CERN Council",
    address = "Geneva",
    year = "2020"
}

@article{deBlas:2024bmz,
    author = "de Blas, Jorge and others",
    title = "{Focus topics for the ECFA study on Higgs / Top / EW factories}",
    eprint = "2401.07564",
    archivePrefix = "arXiv",
    primaryClass = "hep-ph",
    month = "1",
    year = "2024"
}

@article{Bahl:2022igd,
    author = {Bahl, Henning and Biek{\"o}tter, Thomas and Heinemeyer, Sven and Li, Cheng and Paasch, Steven and Weiglein, Georg and Wittbrodt, Jonas},
    title = "{HiggsTools: BSM scalar phenomenology with new versions of HiggsBounds and HiggsSignals}",
    eprint = "2210.09332",
    archivePrefix = "arXiv",
    primaryClass = "hep-ph",
    doi = "10.1016/j.cpc.2023.108803",
    journal = "Comput. Phys. Commun.",
    volume = "291",
    pages = "108803",
    year = "2023"
}

@article{Bernon:2015hsa,
    author = "Bernon, Jeremy and Dumont, Beranger",
    title = "{Lilith: a tool for constraining new physics from Higgs measurements}",
    eprint = "1502.04138",
    archivePrefix = "arXiv",
    primaryClass = "hep-ph",
    doi = "10.1140/epjc/s10052-015-3645-9",
    journal = "Eur. Phys. J. C",
    volume = "75",
    number = "9",
    pages = "440",
    year = "2015"
}

@article{Kraml:2019sis,
    author = "Kraml, Sabine and Loc, Tran Quang and Nhung, Dao Thi and Ninh, Le Duc",
    title = "{Constraining new physics from Higgs measurements with Lilith: update to LHC Run 2 results}",
    eprint = "1908.03952",
    archivePrefix = "arXiv",
    primaryClass = "hep-ph",
    reportNumber = "IFIRSE-TH-2019-5",
    doi = "10.21468/SciPostPhys.7.4.052",
    journal = "SciPost Phys.",
    volume = "7",
    number = "4",
    pages = "052",
    year = "2019"
}

@article{Bertrand:2020lyb,
    author = "Bertrand, Marius and Kraml, Sabine and Loc, Tran Quang and Nhung, Dao Thi and Ninh, Le Duc",
    title = "{Constraining new physics from Higgs measurements with Lilith-2}",
    eprint = "2012.11408",
    archivePrefix = "arXiv",
    primaryClass = "hep-ph",
    reportNumber = "IFIRSE-TH-2020-3",
    doi = "10.22323/1.392.0040",
    journal = "PoS",
    volume = "TOOLS2020",
    pages = "040",
    year = "2021"
}

@article{Athron:2014yba,
    author = {Athron, Peter and Park, Jae-hyeon and St{\"o}ckinger, Dominik and Voigt, Alexander},
    title = "{FlexibleSUSY{\textemdash}A spectrum generator generator for supersymmetric models}",
    eprint = "1406.2319",
    archivePrefix = "arXiv",
    primaryClass = "hep-ph",
    reportNumber = "FTUV-14-3904, IFIC-14-40",
    doi = "10.1016/j.cpc.2014.12.020",
    journal = "Comput. Phys. Commun.",
    volume = "190",
    pages = "139--172",
    year = "2015"
}

@article{Athron:2017fvs,
    author = {Athron, Peter and Bach, Markus and Harries, Dylan and Kwasnitza, Thomas and Park, Jae-hyeon and St{\"o}ckinger, Dominik and Voigt, Alexander and Ziebell, Jobst},
    title = "{FlexibleSUSY 2.0: Extensions to investigate the phenomenology of SUSY and non-SUSY models}",
    eprint = "1710.03760",
    archivePrefix = "arXiv",
    primaryClass = "hep-ph",
    reportNumber = "COEPP-MN-17-16, CoEPP-MN-17-16, KIAS-Q17043, TTK-17-31",
    doi = "10.1016/j.cpc.2018.04.016",
    journal = "Comput. Phys. Commun.",
    volume = "230",
    pages = "145--217",
    year = "2018"
}

@article{Staub:2013tta,
    author = "Staub, Florian",
    title = "{SARAH 4 : A tool for (not only SUSY) model builders}",
    eprint = "1309.7223",
    archivePrefix = "arXiv",
    primaryClass = "hep-ph",
    reportNumber = "BONN-TH-2013-17",
    doi = "10.1016/j.cpc.2014.02.018",
    journal = "Comput. Phys. Commun.",
    volume = "185",
    pages = "1773--1790",
    year = "2014"
}

@article{Staub:2009bi,
    author = "Staub, Florian",
    title = "{From Superpotential to Model Files for FeynArts and CalcHep/CompHep}",
    eprint = "0909.2863",
    archivePrefix = "arXiv",
    primaryClass = "hep-ph",
    doi = "10.1016/j.cpc.2010.01.011",
    journal = "Comput. Phys. Commun.",
    volume = "181",
    pages = "1077--1086",
    year = "2010"
}

@article{Staub:2010jh,
    author = "Staub, Florian",
    title = "{Automatic Calculation of supersymmetric Renormalization Group Equations and Self Energies}",
    eprint = "1002.0840",
    archivePrefix = "arXiv",
    primaryClass = "hep-ph",
    doi = "10.1016/j.cpc.2010.11.030",
    journal = "Comput. Phys. Commun.",
    volume = "182",
    pages = "808--833",
    year = "2011"
}

@article{Staub:2012pb,
    author = "Staub, Florian",
    title = "{SARAH 3.2: Dirac Gauginos, UFO output, and more}",
    eprint = "1207.0906",
    archivePrefix = "arXiv",
    primaryClass = "hep-ph",
    reportNumber = "BONN-TH-2012-17",
    doi = "10.1016/j.cpc.2013.02.019",
    journal = "Comput. Phys. Commun.",
    volume = "184",
    pages = "1792--1809",
    year = "2013"
}

@article{Allanach:2001kg,
    author = "Allanach, B. C.",
    title = "{SOFTSUSY: a program for calculating supersymmetric spectra}",
    eprint = "hep-ph/0104145",
    archivePrefix = "arXiv",
    reportNumber = "CERN-TH-2001-102",
    doi = "10.1016/S0010-4655(01)00460-X",
    journal = "Comput. Phys. Commun.",
    volume = "143",
    pages = "305--331",
    year = "2002"
}

@article{Allanach:2013kza,
    author = "Allanach, B. C. and Athron, P. and Tunstall, Lewis C. and Voigt, A. and Williams, A. G.",
    title = "{Next-to-Minimal SOFTSUSY}",
    eprint = "1311.7659",
    archivePrefix = "arXiv",
    primaryClass = "hep-ph",
    reportNumber = "ADP-13-33-T853, ADP-13-33/T853",
    doi = "10.1016/j.cpc.2014.04.015",
    journal = "Comput. Phys. Commun.",
    volume = "185",
    pages = "2322--2339",
    year = "2014",
    note = "[Erratum: Comput.Phys.Commun. 250, 107044 (2020)]"
}

@article{Athron:2016fuq,
    author = {Athron, Peter and Park, Jae-hyeon and Steudtner, Tom and St{\"o}ckinger, Dominik and Voigt, Alexander},
    title = "{Precise Higgs mass calculations in (non-)minimal supersymmetry at both high and low scales}",
    eprint = "1609.00371",
    archivePrefix = "arXiv",
    primaryClass = "hep-ph",
    reportNumber = "COEPP-MN-16-20, DESY-16-057, KIAS-Q16008",
    doi = "10.1007/JHEP01(2017)079",
    journal = "JHEP",
    volume = "01",
    pages = "079",
    year = "2017"
}

@article{Kwasnitza:2020wli,
    author = {Kwasnitza, Thomas and St{\"o}ckinger, Dominik and Voigt, Alexander},
    title = "{Improved MSSM Higgs mass calculation using the 3-loop FlexibleEFTHiggs approach including $x_{t}$-resummation}",
    eprint = "2003.04639",
    archivePrefix = "arXiv",
    primaryClass = "hep-ph",
    doi = "10.1007/JHEP06(2023)201",
    journal = "JHEP",
    volume = "07",
    number = "07",
    pages = "197",
    year = "2020"
}

@article{Kwasnitza:2025mge,
    author = {Kwasnitza, Thomas and St{\"o}ckinger, Dominik and Voigt, Alexander and W{\"u}nsche, Johannes},
    title = "{Application of the 3-Loop FlexibleEFTHiggs Method to the MSSM and the NMSSM}",
    eprint = "2506.22208",
    archivePrefix = "arXiv",
    primaryClass = "hep-ph",
    month = "6",
    year = "2025"
}

@article{Khasianevich:2024hpv,
    author = {Khasianevich, Uladzimir and Kotlarski, Wojciech and St{\"o}ckinger, Dominik and Voigt, Alexander},
    title = "{FlexibleSUSY extended to automatically compute physical quantities in any beyond the standard model theory: Charged lepton flavor violation processes, Higgs decays, and user-defined observables}",
    eprint = "2402.14630",
    archivePrefix = "arXiv",
    primaryClass = "hep-ph",
    doi = "10.1016/j.cpc.2024.109244",
    journal = "Comput. Phys. Commun.",
    volume = "302",
    pages = "109244",
    year = "2024"
}

@article{Athron:2022isz,
    author = {Athron, Peter and Bach, Markus and Jacob, Douglas H. J. and Kotlarski, Wojciech and St{\"o}ckinger, Dominik and Voigt, Alexander},
    title = "{Precise calculation of the W boson pole mass beyond the standard model with FlexibleSUSY}",
    eprint = "2204.05285",
    archivePrefix = "arXiv",
    primaryClass = "hep-ph",
    doi = "10.1103/PhysRevD.106.095023",
    journal = "Phys. Rev. D",
    volume = "106",
    number = "9",
    pages = "095023",
    year = "2022"
}

@article{Athron:2021kve,
    author = {Athron, Peter and B{\"u}chner, Adam and Harries, Dylan and Kotlarski, Wojciech and St{\"o}ckinger, Dominik and Voigt, Alexander},
    title = "{FlexibleDecay: An automated calculator of scalar decay widths}",
    eprint = "2106.05038",
    archivePrefix = "arXiv",
    primaryClass = "hep-ph",
    doi = "10.1016/j.cpc.2022.108584",
    journal = "Comput. Phys. Commun.",
    volume = "283",
    pages = "108584",
    year = "2023"
}

@article{Kotlarski:2019muo,
    author = {Kotlarski, Wojciech and St{\"o}ckinger, Dominik and St{\"o}ckinger-Kim, Hyejung},
    title = "{Low-energy lepton physics in the MRSSM: $(g-2)_\mu$, $\mu \to e\gamma$ and $\mu\to e$ conversion}",
    eprint = "1902.06650",
    archivePrefix = "arXiv",
    primaryClass = "hep-ph",
    doi = "10.1007/JHEP08(2019)082",
    journal = "JHEP",
    volume = "08",
    pages = "082",
    year = "2019"
}

@article{Dudenas:2022von,
    author = {D{\={u}}d{\.{e}}nas, Vytautas and Gajdosik, Thomas and Khasianevich, Uladzimir and Kotlarski, Wojciech and St{\"o}ckinger, Dominik},
    title = "{Charged lepton flavor violating processes in the Grimus-Neufeld model}",
    eprint = "2206.00661",
    archivePrefix = "arXiv",
    primaryClass = "hep-ph",
    doi = "10.1007/JHEP09(2022)174",
    journal = "JHEP",
    volume = "09",
    pages = "174",
    year = "2022"
}

@article{Dudenas:2022xnq,
    author = {D{\={u}}d{\.{e}}nas, Vytautas and Gajdosik, Thomas and Khasianevich, Uladzimir and Kotlarski, Wojciech and St{\"o}ckinger, Dominik},
    title = "{Box-enhanced charged lepton flavor violation in the Grimus-Neufeld model}",
    eprint = "2211.14384",
    archivePrefix = "arXiv",
    primaryClass = "hep-ph",
    doi = "10.1103/PhysRevD.107.055027",
    journal = "Phys. Rev. D",
    volume = "107",
    number = "5",
    pages = "055027",
    year = "2023"
}

@article{Athron:2021iuf,
    author = {Athron, Peter and Bal{\'a}zs, Csaba and Jacob, Douglas H. J. and Kotlarski, Wojciech and St{\"o}ckinger, Dominik and St{\"o}ckinger-Kim, Hyejung},
    title = "{New physics explanations of a$_{\mu}$ in light of the FNAL muon g {\ensuremath{-}} 2 measurement}",
    eprint = "2104.03691",
    archivePrefix = "arXiv",
    primaryClass = "hep-ph",
    doi = "10.1007/JHEP09(2021)080",
    journal = "JHEP",
    volume = "09",
    pages = "080",
    year = "2021"
}

@article{Kalinowski:2024uxe,
    author = "Kalinowski, Jan and Kotlarski, Wojciech",
    title = "{Interpreting 95 GeV di-photon/$ b\overline{b} $ excesses as a lightest Higgs boson of the MRSSM}",
    eprint = "2403.08720",
    archivePrefix = "arXiv",
    primaryClass = "hep-ph",
    doi = "10.1007/JHEP07(2024)037",
    journal = "JHEP",
    volume = "07",
    pages = "037",
    year = "2024"
}

@article{Kotlarski:2025jvf,
    author = "Kotlarski, Wojciech and Patellis, Gregory",
    title = "{Phenomenology of the Higgs sector from Reduction of Couplings in the Type-II 2HDM}",
    eprint = "2502.20357",
    archivePrefix = "arXiv",
    primaryClass = "hep-ph",
    month = "2",
    year = "2025"
}

@article{Hahn:1998yk,
    author = "Hahn, T. and Perez-Victoria, M.",
    title = "{Automatized one loop calculations in four-dimensions and D-dimensions}",
    eprint = "hep-ph/9807565",
    archivePrefix = "arXiv",
    reportNumber = "UG-FT-87-98, KA-TP-7-1998",
    doi = "10.1016/S0010-4655(98)00173-8",
    journal = "Comput. Phys. Commun.",
    volume = "118",
    pages = "153--165",
    year = "1999"
}

@article{Denner:2002ii,
    author = "Denner, Ansgar and Dittmaier, S.",
    title = "{Reduction of one loop tensor five point integrals}",
    eprint = "hep-ph/0212259",
    archivePrefix = "arXiv",
    reportNumber = "MPI-PHT-2002-63, PSI-PR-02-21",
    doi = "10.1016/S0550-3213(03)00184-6",
    journal = "Nucl. Phys. B",
    volume = "658",
    pages = "175--202",
    year = "2003"
}

@article{Denner:2005nn,
    author = "Denner, Ansgar and Dittmaier, S.",
    title = "{Reduction schemes for one-loop tensor integrals}",
    eprint = "hep-ph/0509141",
    archivePrefix = "arXiv",
    reportNumber = "MPP-2005-84, PSI-PR-05-08",
    doi = "10.1016/j.nuclphysb.2005.11.007",
    journal = "Nucl. Phys. B",
    volume = "734",
    pages = "62--115",
    year = "2006"
}

@article{Denner:2010tr,
    author = "Denner, A. and Dittmaier, S.",
    title = "{Scalar one-loop 4-point integrals}",
    eprint = "1005.2076",
    archivePrefix = "arXiv",
    primaryClass = "hep-ph",
    reportNumber = "FR-PHENO-2010-020, PSI-PR-10-10",
    doi = "10.1016/j.nuclphysb.2010.11.002",
    journal = "Nucl. Phys. B",
    volume = "844",
    pages = "199--242",
    year = "2011"
}

@article{Denner:2016kdg,
    author = "Denner, Ansgar and Dittmaier, Stefan and Hofer, Lars",
    title = "{Collier: a fortran-based Complex One-Loop LIbrary in Extended Regularizations}",
    eprint = "1604.06792",
    archivePrefix = "arXiv",
    primaryClass = "hep-ph",
    reportNumber = "FR-PHENO-2016-003, ICCUB-16-016",
    doi = "10.1016/j.cpc.2016.10.013",
    journal = "Comput. Phys. Commun.",
    volume = "212",
    pages = "220--238",
    year = "2017"
}

@article{Skands:2003cj,
    author = "Skands, Peter Z. and others",
    title = "{SUSY Les Houches accord: Interfacing SUSY spectrum calculators, decay packages, and event generators}",
    eprint = "hep-ph/0311123",
    archivePrefix = "arXiv",
    reportNumber = "LU-TP-03-39, SHEP-03-24, CERN-TH-2003-204, ZU-TH-15-03, LMU-19-03, DCPT-03-108, IPPP-03-54, CTS-IISC-2003-07, DESY-03-166, MPP-2003-111",
    doi = "10.1088/1126-6708/2004/07/036",
    journal = "JHEP",
    volume = "07",
    pages = "036",
    year = "2004"
}

@article{Allanach:2008qq,
    author = "Allanach, B. C. and others",
    title = "{SUSY Les Houches Accord 2}",
    eprint = "0801.0045",
    archivePrefix = "arXiv",
    primaryClass = "hep-ph",
    reportNumber = "FERMILAB-PUB-07-036-T, SLAC-PUB-12765, CERN-PH-TH-2007-148, DAMTP-2007-76, EDINBURGH-2007-31, KEK-TH-1170, LAPTH-1204-07, LPT-ORSAY-07-81, SHEP-07-13",
    doi = "10.1016/j.cpc.2008.08.004",
    journal = "Comput. Phys. Commun.",
    volume = "180",
    pages = "8--25",
    year = "2009"
}

@article{ParticleDataGroup:2024cfk,
    author = "Navas, S. and others",
    collaboration = "Particle Data Group",
    title = "{Review of particle physics}",
    doi = "10.1103/PhysRevD.110.030001",
    journal = "Phys. Rev. D",
    volume = "110",
    number = "3",
    pages = "030001",
    year = "2024"
}

@article{Biekotter:2023oen,
    author = {Biek{\"o}tter, Thomas and Heinemeyer, Sven and Weiglein, Georg},
    title = "{95.4~GeV diphoton excess at ATLAS and CMS}",
    eprint = "2306.03889",
    archivePrefix = "arXiv",
    primaryClass = "hep-ph",
    reportNumber = "KA-TP-11-2023, DESY-23-071, IFT--UAM/CSIC-23-062",
    doi = "10.1103/PhysRevD.109.035005",
    journal = "Phys. Rev. D",
    volume = "109",
    number = "3",
    pages = "035005",
    year = "2024"
}

@article{LEPWorkingGroupforHiggsbosonsearches:2003ing,
    author = "Barate, R. and others",
    collaboration = "LEP Working Group for Higgs boson searches, ALEPH, DELPHI, L3, OPAL",
    title = "{Search for the standard model Higgs boson at LEP}",
    eprint = "hep-ex/0306033",
    archivePrefix = "arXiv",
    reportNumber = "CERN-EP-2003-011",
    doi = "10.1016/S0370-2693(03)00614-2",
    journal = "Phys. Lett. B",
    volume = "565",
    pages = "61--75",
    year = "2003"
}

@article{CMS:2024yhz,
    author = "Hayrapetyan, Aram and others",
    collaboration = "CMS",
    title = "{Search for a standard model-like Higgs boson in the mass range between 70 and 110 GeV in the diphoton final state in proton-proton collisions at $\sqrt{s}=13$~TeV}",
    eprint = "2405.18149",
    archivePrefix = "arXiv",
    primaryClass = "hep-ex",
    reportNumber = "CMS-HIG-20-002, CERN-EP-2024-088",
    doi = "10.1016/j.physletb.2024.139067",
    journal = "Phys. Lett. B",
    volume = "860",
    pages = "139067",
    year = "2025"
}

@article{ATLAS:2020tlo,
    author = "Aad, Georges and others",
    collaboration = "ATLAS",
    title = "{Search for heavy resonances decaying into a pair of Z bosons in the $\ell ^+\ell ^-\ell '^+\ell '^-$ and $\ell ^+\ell ^-\nu {{\bar{\nu }}}$ final states using 139 $\mathrm {fb}^{-1}$ of proton{\textendash}proton collisions at $\sqrt{s} = 13\,$TeV with the ATLAS detector}",
    eprint = "2009.14791",
    archivePrefix = "arXiv",
    primaryClass = "hep-ex",
    reportNumber = "CERN-EP-2020-153",
    doi = "10.1140/epjc/s10052-021-09013-y",
    journal = "Eur. Phys. J. C",
    volume = "81",
    number = "4",
    pages = "332",
    year = "2021"
}

@article{CMS:2022goy,
    author = "Tumasyan, Armen and others",
    collaboration = "CMS",
    title = "{Searches for additional Higgs bosons and for vector leptoquarks in $\tau\tau$ final states in proton-proton collisions at $\sqrt{s}$ = 13 TeV}",
    eprint = "2208.02717",
    archivePrefix = "arXiv",
    primaryClass = "hep-ex",
    reportNumber = "CMS-HIG-21-001, CERN-EP-2022-137",
    doi = "10.1007/JHEP07(2023)073",
    journal = "JHEP",
    volume = "07",
    pages = "073",
    year = "2023"
}

@article{ATLAS:2022akr,
    author = "Aad, Georges and others",
    collaboration = "ATLAS",
    title = "{Measurement of the CP properties of Higgs boson interactions with $\tau $-leptons with the ATLAS detector}",
    eprint = "2212.05833",
    archivePrefix = "arXiv",
    primaryClass = "hep-ex",
    reportNumber = "CERN-EP-2022-244",
    doi = "10.1140/epjc/s10052-023-11583-y",
    journal = "Eur. Phys. J. C",
    volume = "83",
    number = "7",
    pages = "563",
    year = "2023"
}

@article{CMS:2021sdq,
    author = "Tumasyan, Armen and others",
    collaboration = "CMS",
    title = "{Analysis of the $CP$ structure of the Yukawa coupling between the Higgs boson and $\tau$ leptons in proton-proton collisions at $ \sqrt{s} $ = 13 TeV}",
    eprint = "2110.04836",
    archivePrefix = "arXiv",
    primaryClass = "hep-ex",
    reportNumber = "CMS-HIG-20-006, CERN-EP-2021-189",
    doi = "10.1007/JHEP06(2022)012",
    journal = "JHEP",
    volume = "06",
    pages = "012",
    year = "2022"
}

@article{ATLAS:2023cbt,
    author = "Aad, Georges and others",
    collaboration = "ATLAS",
    title = "{Probing the CP nature of the top{\textendash}Higgs Yukawa coupling in tt{\textasciimacron}H and tH events with H{\textrightarrow}bb{\textasciimacron} decays using the ATLAS detector at the LHC}",
    eprint = "2303.05974",
    archivePrefix = "arXiv",
    primaryClass = "hep-ex",
    reportNumber = "CERN-EP-2022-208",
    doi = "10.1016/j.physletb.2024.138469",
    journal = "Phys. Lett. B",
    volume = "849",
    pages = "138469",
    year = "2024"
}

@article{ATLAS:2023tkt,
    author = "Aad, Georges and others",
    collaboration = "ATLAS",
    title = "{Combination of searches for invisible decays of the Higgs boson using 139 fb{\ensuremath{-}}1 of proton-proton collision data at s=13 TeV collected with the ATLAS experiment}",
    eprint = "2301.10731",
    archivePrefix = "arXiv",
    primaryClass = "hep-ex",
    reportNumber = "CERN-EP-2022-289",
    doi = "10.1016/j.physletb.2023.137963",
    journal = "Phys. Lett. B",
    volume = "842",
    pages = "137963",
    year = "2023"
}

@article{CMS:2023sdw,
    author = "Tumasyan, Armen and others",
    collaboration = "CMS",
    title = "{A search for decays of the Higgs boson to invisible particles in events with a top-antitop quark pair or a vector boson in proton-proton collisions at $\sqrt{s} = 13\,\text {Te}\hspace{-.08em}\text {V} $}",
    eprint = "2303.01214",
    archivePrefix = "arXiv",
    primaryClass = "hep-ex",
    reportNumber = "CMS-HIG-21-007, CERN-EP-2023-004",
    doi = "10.1140/epjc/s10052-023-11952-7",
    journal = "Eur. Phys. J. C",
    volume = "83",
    number = "10",
    pages = "933",
    year = "2023"
}

@article{Kribs:2007ac,
    author = "Kribs, Graham D. and Poppitz, Erich and Weiner, Neal",
    title = "{Flavor in supersymmetry with an extended R-symmetry}",
    eprint = "0712.2039",
    archivePrefix = "arXiv",
    primaryClass = "hep-ph",
    doi = "10.1103/PhysRevD.78.055010",
    journal = "Phys. Rev. D",
    volume = "78",
    pages = "055010",
    year = "2008"
}

@article{Diessner:2015iln,
    author = {Diessner, Philip and Kalinowski, Jan and Kotlarski, Wojciech and St{\"o}ckinger, Dominik},
    title = "{Exploring the Higgs sector of the MRSSM with a light scalar}",
    eprint = "1511.09334",
    archivePrefix = "arXiv",
    primaryClass = "hep-ph",
    doi = "10.1007/JHEP03(2016)007",
    journal = "JHEP",
    volume = "03",
    pages = "007",
    year = "2016"
}

@article{Fayet:1974pd,
    author = "Fayet, Pierre",
    title = "{Supergauge Invariant Extension of the Higgs Mechanism and a Model for the electron and Its Neutrino}",
    reportNumber = "PTENS-74-7",
    doi = "10.1016/0550-3213(75)90636-7",
    journal = "Nucl. Phys. B",
    volume = "90",
    pages = "104--124",
    year = "1975"
}

@article{Diessner:2014ksa,
    author = {Die{\ss}ner, Philip and Kalinowski, Jan and Kotlarski, Wojciech and St{\"o}ckinger, Dominik},
    title = "{Higgs boson mass and electroweak observables in the MRSSM}",
    eprint = "1410.4791",
    archivePrefix = "arXiv",
    primaryClass = "hep-ph",
    doi = "10.1007/JHEP12(2014)124",
    journal = "JHEP",
    volume = "12",
    pages = "124",
    year = "2014"
}

@article{ATLAS:2023dnm,
    author = "Aad, Georges and others",
    collaboration = "ATLAS",
    title = "{Evidence of off-shell Higgs boson production from ZZ leptonic decay channels and constraints on its total width with the ATLAS detector}",
    eprint = "2304.01532",
    archivePrefix = "arXiv",
    primaryClass = "hep-ex",
    reportNumber = "CERN-EP-2023-023",
    doi = "10.1016/j.physletb.2024.138734",
    journal = "Phys. Lett. B",
    volume = "846",
    pages = "138223",
    year = "2023",
    note = "[Erratum: Phys.Lett.B 854, 138734 (2024), Erratum: Phys.Lett.B 865, 139449 (2025)]"
}

@article{CMS-PAS-HIG-24-011,
    collaboration = "CMS",
    title = "{Measurement of the Higgs boson decay width from off-shell production using the WW decay in the all-leptonic final state in pp collisions at 13 TeV}",
    reportNumber = "CMS-PAS-HIG-24-011",
    year = "2025"
}

@article{deBlas:2019rxi,
    author = "de Blas, J. and others",
    title = "{Higgs Boson Studies at Future Particle Colliders}",
    eprint = "1905.03764",
    archivePrefix = "arXiv",
    primaryClass = "hep-ph",
    reportNumber = "DESY-19-079",
    doi = "10.1007/JHEP01(2020)139",
    journal = "JHEP",
    volume = "01",
    pages = "139",
    year = "2020"
}

@article{Lang:2025eom,
    author = "Lang, Jonas and Kotlarski, Wojciech",
    title = "{Towards Automatizing Higgs Decays in BSM Models at One-loop in the Decoupling Renormalization Scheme}",
    eprint = "2504.13577",
    archivePrefix = "arXiv",
    primaryClass = "hep-ph",
    doi = "10.5506/APhysPolBSupp.18.5-A18",
    journal = "Acta Phys. Polon. Supp.",
    volume = "18",
    number = "5",
    pages = "5-A18",
    year = "2025"
}

@article{Goodsell:2015ira,
    author = "Goodsell, M. and Nickel, K. and Staub, F.",
    title = "{Generic two-loop Higgs mass calculation from a diagrammatic approach}",
    eprint = "1503.03098",
    archivePrefix = "arXiv",
    primaryClass = "hep-ph",
    reportNumber = "BONN-TH-2015-04, CERN-PH-TH-2015-044",
    doi = "10.1140/epjc/s10052-015-3494-6",
    journal = "Eur. Phys. J. C",
    volume = "75",
    number = "6",
    pages = "290",
    year = "2015"
}

@article{harris2020array,
 title         = {Array programming with {NumPy}},
 author        = {Charles R. Harris and K. Jarrod Millman and St{\'{e}}fan J.
                 van der Walt and Ralf Gommers and Pauli Virtanen and David
                 Cournapeau and Eric Wieser and Julian Taylor and Sebastian
                 Berg and Nathaniel J. Smith and Robert Kern and Matti Picus
                 and Stephan Hoyer and Marten H. van Kerkwijk and Matthew
                 Brett and Allan Haldane and Jaime Fern{\'{a}}ndez del
                 R{\'{i}}o and Mark Wiebe and Pearu Peterson and Pierre
                 G{\'{e}}rard-Marchant and Kevin Sheppard and Tyler Reddy and
                 Warren Weckesser and Hameer Abbasi and Christoph Gohlke and
                 Travis E. Oliphant},
 year          = {2020},
 month         = sep,
 journal       = {Nature},
 volume        = {585},
 number        = {7825},
 pages         = {357--362},
 doi           = {10.1038/s41586-020-2649-2},
 publisher     = {Springer Science and Business Media {LLC}},
 url           = {https://doi.org/10.1038/s41586-020-2649-2}
}
\addcontentsline{toc}{section}{References}

\end{document}